\pdfoutput=1
\documentclass{emulateapj}
\usepackage{graphics,graphicx}
\usepackage{fancyheadings}
\usepackage{color}
\usepackage{natbib}
\usepackage{times}

\newcommand{\suz}{\textit{Suzaku}}
\newcommand{\wise}{\textit{WISE}}
\newcommand{\lat}{\textit{Fermi}-LAT}
\newcommand{\cen}{Centaurus\,A}
\def\as{\hbox{$^{\prime\prime}$}}
\def\am{\hbox{$^{\prime}$}}
\def\deg{\hbox{$^\circ$}}
\newcommand{\Lu}{erg\,s$^{-1}$}
\newcommand{\Su}{erg\,cm$^{-2}$\,s$^{-1}$}
\newcommand{\Pu}{ph\,cm$^{-2}$\,s$^{-1}$}
\newcommand{\Uu}{erg\,cm$^{-3}$}
\newcommand{\bmg}{B_{\rm \mu G}}

\shorttitle{Giant Lobes of Cen\,A in X-rays}
\shortauthors{Stawarz et al.}

\begin{document}

\title{Giant Lobes of Centaurus\,A Radio Galaxy Observed with the Suzaku X-ray Satellite}

\author{\L .~Stawarz$^{1,\,2}$, Y.~T.~Tanaka$^3$, G.~Madejski$^4$, S.~P.~O'Sullivan$^5$, C.~C.~Cheung$^6$, I.~J.~Feain$^7$, Y.~Fukazawa$^3$, P.~Gandhi$^1$, M.~J.~Hardcastle$^8$, J.~Kataoka$^9$, M.~Ostrowski$^2$, B.~Reville$^{10}$, A.~Siemiginowska$^{11}$, A.~Simionescu$^{12}$, T.~Takahashi$^1$, Y.~Takei$^1$, Y.~Takeuchi$^9$, and N.~Werner$^{12}$}

\medskip

\affil{$^1$ Institute of Space and Astronautical Science JAXA, 3-1-1 Yoshinodai, Chuo-ku, Sagamihara, Kanagawa 252-5210, Japan} 
\affil{$^2$ Astronomical Observatory, Jagiellonian University, ul. Orla 171, 30-244 Krak\'ow, Poland}
\affil{$^3$ Department of Physical Sciences, Hiroshima University, Higashi-Hiroshima, Hiroshima 739-8526, Japan}
\affil{$^4$ W. W. Hansen Experimental Physics Laboratory, Kavli Institute for Particle Astrophysics and Cosmology, Department of Physics and SLAC National Accelerator Laboratory, Stanford University, Stanford, CA 94305, USA}
\affil{$^5$ Sydney Institute for Astronomy, School of Physics A28, University of Sydney, NSW 2006, Australia}
\affil{$^6$ National Research Council Research Associate, National Academy of Sciences, Washington, DC 20001, resident at Naval Research Laboratory, Washington, DC 20375, USA}
\affil{$^7$ CSIRO Astronomy and Space Science, ATNF, PO Box 76, Epping, NSW 1710, Australia}
\affil{$^8$ School of Physics, Astronomy \& Mathematics, University of Hertfordshire, College Lane, Hatfield AL10 9AB, UK}
\affil{$^9$ Research Institute for Science and Engineering, Waseda University, 3-4-1, Okubo, Shinjuku, Tokyo 169-8555, Japan}
\affil{$^{10}$ Clarendon Laboratory, University of Oxford, Parks Road, Oxford OX1 3PU, UK}
\affil{$^{11}$ Harvard Smithsonian Center for Astrophysics, 60 Garden St, Cambridge, MA 02138, USA}
\affil{$^{12}$ KIPAC, Stanford University, 452 Lomita Mall, Stanford, CA 94305, USA, and Department of Physics, Stanford University, 382 Via Pueblo Mall, Stanford, CA 94305-4060, USA}

\medskip

\email{stawarz@astro.isas.jaxa.jp}
\label{firstpage}

\begin{abstract}

We report on \suz\ observations of selected regions within the
Southern giant lobe of the radio galaxy \cen. In our analysis we focus
on distinct X-ray features detected with the X-ray Imaging
Spectrometer within the range $0.5-10$\,keV, some of which are
\emph{likely} associated with fine structure of the lobe revealed by
recent high-quality radio intensity and polarization maps. With the
available photon statistics, we find that the spectral properties of
the detected X-ray features are equally consistent with thermal
emission from hot gas with temperatures $kT > 1$\,keV, or with a
power-law radiation continuum characterized by photon indices $\Gamma
\sim 2.0 \pm 0.5$. However, the plasma parameters implied by these different
models favor a synchrotron origin for the analyzed X-ray spots,
indicating that a very efficient acceleration of electrons up to
$\gtrsim 10$\,TeV energies is taking place within the giant structure
of \cen, albeit only in isolated and compact regions associated with
extended and highly polarized radio filaments. We also present a
detailed analysis of the diffuse X-ray emission filling the whole
field-of-view of the instrument, resulting in a \emph{tentative}
detection of a soft excess component best fitted by a thermal model
with a temperature of $kT \sim 0.5$\,keV. The exact origin of the
observed excess remains uncertain, although energetic considerations
point to thermal gas filling the bulk of the volume of the lobe and
mixed with the non-thermal plasma, rather than to the alternative
scenario involving a condensation of the hot intergalactic medium
around the edges of the expanding radio structure. If correct, this
would be the first detection of the thermal content of the extended
lobes of a radio galaxy in X-rays. The corresponding number density of
the thermal gas in such a case is $n_g \sim 10^{-4}$\,cm$^{-3}$,
while its pressure appears to be in almost exact equipartition with
the volume-averaged non-thermal pressure provided by the
radio-emitting electrons and the lobes' magnetic field. A prominent
large-scale fluctuation of the Galactic foreground emission, resulting
in excess foreground X-ray emission aligned with the lobe, cannot be
ruled out. Although tentative, our findings potentially imply that the
structure of the extended lobes in active galaxies is likely to be
highly inhomogeneous and non-uniform, with magnetic reconnection and
turbulent acceleration processes continuously converting magnetic
energy to internal energy of the plasma particles, leading to
possibly significant spatial and temporal variations in the plasma
$\beta$ parameter around the volume-averaged equilibrium condition
$\beta \sim 1$.
\end{abstract}


\section{Introduction}
\label{sec:intro}

The radio source \cen\ is hosted by a massive elliptical galaxy
NGC\,5128, located at a distance of $D=3.7$\,Mpc \citep[for a review
  see][]{Israel98}. As such, it is the closest active galaxy, and its
innermost structure, down to hundreds of Schwarzschild radii of the
central black hole \citep[mass $M_{\rm BH} \simeq 10^8
  M_{\odot}$;][]{Neumayer10}, can be studied in great detail with
modern high-resolution and high-sensitivity instruments. Yet \cen\ is
at the same time surrounded by a giant radio halo, or rather giant
lobes, extending for about $\sim 600$\,kpc in the North-South
direction and $\sim 200$ kpc wide. This fossil structure, although it
has been known for a while \citep{Cooper65}, is still relatively
poorly studied, because of its huge angular size
($\sim$\,5\,deg\,$\times$\,9\,deg), which in the past has precluded
detailed observational studies. This situation has
changed only very recently, and the giant lobes of the \cen\ system
can now be investigated
with the improved angular resolution available at radio and $\gamma$-ray
frequencies.

Since its discovery, the giant halo of \cen\ (total radio luminosity
$L_{\rm R} \simeq 10^{41}$\,\Lu) has been observed multiple times with
low-resolution single-dish and space-borne radio telescopes. Detailed
analysis indicated that the radio spectrum of the Southern lobe
steepens systematically both with frequency and with distance from the
core, in agreement with the expectation for an aging relic ($\sim
30$\,Myr-old) system with the \emph{volume-averaged} equipartition
magnetic field $B_{\rm eq} \simeq 1.3$\,$\mu$G. Meanwhile, the radio
spectrum of the Northern lobe was found to be roughly
position-independent, inconsistent with the scenario of passive aging,
but suggesting instead an ongoing or at least recent injection of
radiating electrons throughout the entire volume of the lobe
\citep{Hardcastle09}. The most recent high-resolution radio
observations of the giant structure at 1.4\,GHz with ATCA and Parkes
64m telescopes revealed a very inhomogeneous and filamentary structure
of the lobes \citep{Feain11}. The 50\as-resolution map derived from
this study ---
``the most detailed radio continuum image of any radio galaxy to
date'' -- showed that the Northern lobe constitutes an extension of
the inner structure (down to the radio core), for which the outer part
could be resolved into a series of semi-regularly spaced ($\simeq
25$\,kpc) and concentric shells or filaments with a projected
thickness of $3-6$\,kpc, particularly prominent in the high-resolution
radio polarized intensity maps. By contrast, the Southern lobe, which
appears to be physically unconnected to the core, displayed a largely
chaotic and mottled morphology.

The magnetic field structure and thermal matter content of the giant
lobes in \cen\ were also probed recently at radio frequencies by an
analysis of the Faraday Rotation Measure (RM) using the ensemble of
polarized background sources \citep{Feain09}. The RM analysis showed
that the polarized emission in the Northern lobe follows closely the
continuum emission down to the sensitivity limits of the survey, with
little evidence for depolarization, whereas the emission in
the Southern lobe is depolarized and chaotic with large jumps in the
polarization angle. The apparent lack of \emph{internal}
depolarization effects for most of the giant structure enabled a
meaningful limit to be placed on the \emph{volume-averaged} density of
the thermal plasma \emph{within} the lobes $\bar{n}_g < 0.5 \times
10^{-4} (B/B_{\rm eq})^{-1}$\,cm$^{-3}$. This limit is not inconsistent
with the analysis of the improved RM data presented in the most recent 
paper \citet{Shane12}, resulting in a positive detection 
of the internal depolarization signal, with the corresponding gas density 
being $\bar{n}_g \sim 10^{-4}$\,cm$^{-3}$. The revealed non-uniform
distribution of the RM excess throughout the entire volume of the
source should however be kept in mind, as it reflects a highly inhomogeneous distribution of the thermal
gas mixed with the lobes, and/or a complex topology of the lobes'
magnetic field.

\cen\ is a source of $\gamma$-ray emission within the broad energy
range from $\gtrsim 50$\,keV up to $\lesssim 10$\,TeV, as established
by the pioneering observations with Compton Gamma-Ray Observatory
\citep{Steinle98} and H.E.S.S. Cherenkov Telescope \citep{HESS09}. It
was widely believed, however, that it is the active core (nucleus and
nuclear jet) which dominates the production of $\gamma$-rays in the
system, especially because this core appears as a very prominent X-ray
emitter with a complex, multi-component spectrum \citep[see,
  e.g.,][and references therein]{Fukazawa11}. And, indeed, even the
most recent analysis of the INTEGRAL/SPI data resulted in only upper
limits for the soft $\gamma$-ray emission of the giant lobes
\citep[$0.04-1$\,MeV photon flux $<1.1 \times
  10^{-3}$\,\Pu;][]{Beckmann11}. On the other hand, during the first
10 months of its operation, the new-generation $\gamma$-ray satellite
\lat\ resolved the giant lobes of \cen\ \citep{LAT10}, revealing
that they contribute about half of the total flux of the whole system
at MeV/GeV photon energies (the $0.1-100$\,GeV lobes' luminosity is
$L_{\gamma} \simeq 10^{41}$\Lu). This important result proved that
$\gamma$-rays are being efficiently generated within the extended
halo, despite the advanced age and relaxed nature of the
structure.\footnote{See also in this context \citet{Takeuchi12} and
  \citet{Katsuta12} for the cases of two other radio galaxies with the
  extended lobes possibly resolved or at least detected by \lat.} The
detected emission, well modeled as inverse-Compton up-scattering of
cosmic background photons by ultrarelativistic electrons, moreover implies
that the giant lobes in \cen\ are close to the condition of
energy/pressure equipartition between the radiating particles and the
magnetic field (volume-averaged field strength $\bar{B} \simeq
0.9$\,$\mu$G\,$\lesssim B_{\rm eq}$).

Despite these new data, our understanding of the giant lobes in
\cen\ is still incomplete. Particularly missing in this context is
X-ray information, even though the large-scale structure of \cen\ has
been the subject of low-resolution, low-sensitivity observations with 
several previous X-ray satellites. The first meaningful upper limits
for the X-ray emission of the giant lobes in the system were provided
by the SAS\,3 instrument \citep{Marshall81}, which probed the
$2-10$\,keV continuum of the target down to a flux level of $3\times
10^{-11}$\,\Su\ ($90\%$ limit). Subsequent observations with
\textit{ROSAT} suggested the presence of a giant X-ray structure
around the radio halo \citep{Arp94}, which was, however, not confirmed
by later analysis. Finally, ASCA observations of the outer part of the
Northern lobe revealed the presence of a compact X-ray feature with the
$2-10$\,keV flux $\sim 10^{-12}$\,\Su, which was modeled as a power-law
continuum or equivalently by thermal emission from a very hot gas
\citep[$kT \sim 10$\,keV;][]{Isobe01}. In addition, a soft diffuse
emission component with a total $0.5-2$\,keV flux $\sim 8.5 \times
10^{-14}$\,\Su/0.55\,deg$^{2}$, distributed within the whole
field-of-view of the GIS detector, best fitted by a thermal model with
$kT \sim 0.6$\,keV, was tentatively found. Our preliminary analysis of
the archival ASCA data confirmed the presence of the hard X-ray spot,
consisting of several intensity peaks aligned roughly with a highly
polarized radio filament. However, some of the X-ray emission peaks
coincide with the positions of optical background radio sources, and
as such may be unrelated to the giant lobes. A more detailed
discussion on the archival ASCA data, together with the analysis of the
forthcoming follow-up \suz\ and \textit{Chandra} observations will be
published in a later paper.

To close the gap between the X-ray studies of the giant halo in
\cen\ and the most recent observations at other wavelengths, in
previous years we have requested the very first \suz\ observations of
the Southern lobe (AO-5, $\#$51401). The low and relatively constant instrumental 
background of \suz\ is ideally suited for observations of low-surface brightness diffuse
sources such as giant lobes of \cen. In addition, the spatial resolution of the
\suz's XIS ($\simeq 2$\am) is well matched to the recently obtained
high-resolution (50\as) radio maps of the giant lobes, while the FOV
of the instrument ($\simeq 1/3$\,deg\,$\times 1/3$\,deg) is large
enough to encapsulate within a single \suz\ pointing the large-scale
structures (shells/filaments) revealed by the radio polarization maps;
\suz\ is therefore an ideal instrument for this work. However, because of 
the large extension of the giant lobes on the sky, only a small part of their 
structure could be covered in the two 80-ks exposures that were performed. 
In this paper we report on the analysis of these \suz\ data, focusing firstly on
distinct X-ray features detected within the field-of-view (FOV) of the
X-ray Imaging Spectrometer (XIS) centered on the radio-bright part of
the Southern lobe, and secondly on the diffuse emission filling the whole
FOV of the instrument and apparently related to the \cen\ halo. Our
results are presented in detail in \S\ref{sec:data} and discussed further in
\S\ref{sec:discussion} below.

\section{Suzaku Data}
\label{sec:data}

\subsection{Observations and Data Processing}
\label{sec:analysis}

\begin{figure}[!th]
\begin{center}
\includegraphics[width=5.0in,angle=270]{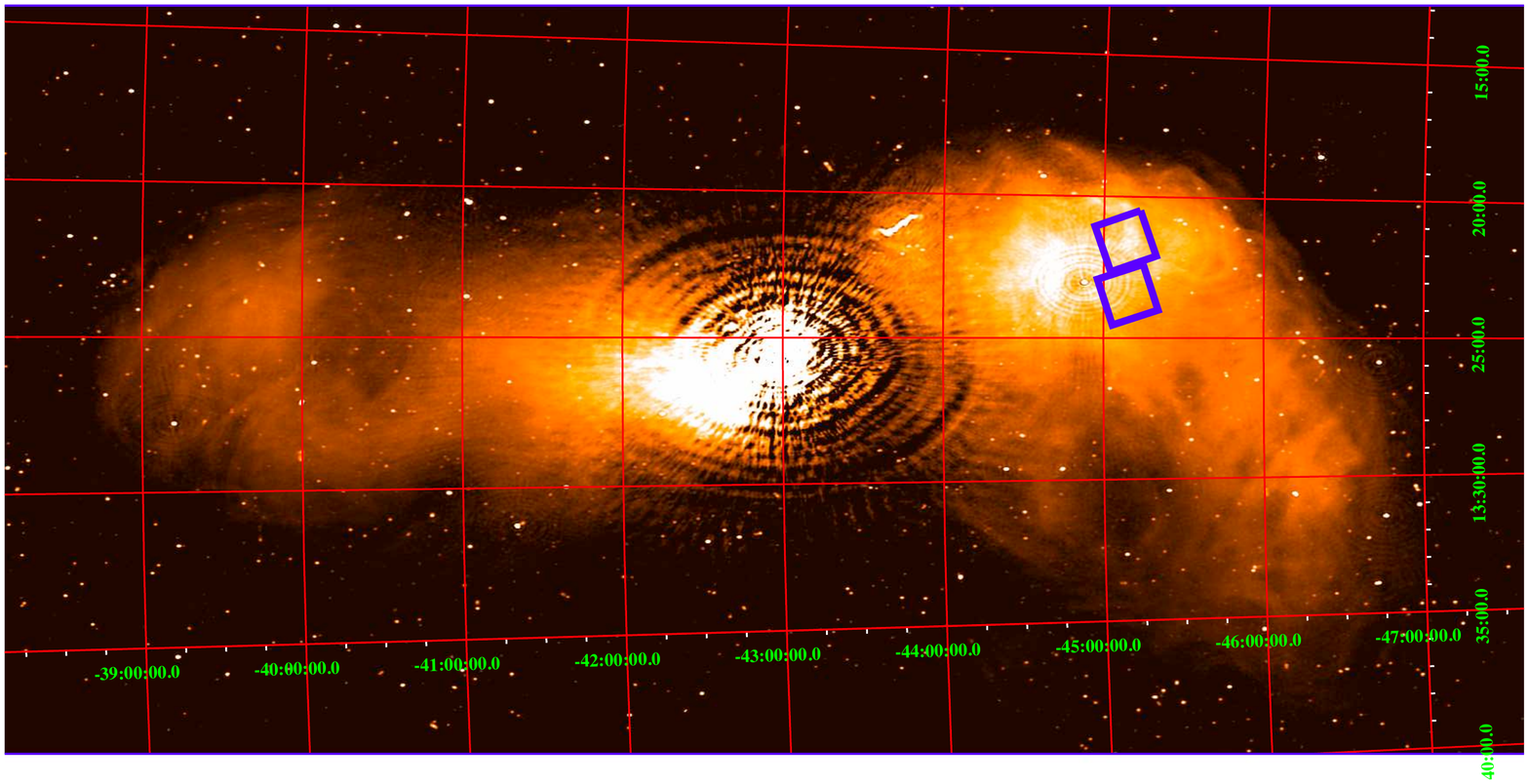}
\caption{\small 1.4\,GHz total intensity map of the giant \cen\ structure \citep{Feain11} with the two \suz\ ``on'' pointings marked as blue squares (see Table\,\ref{tab:obslog}).}
\label{fig:X-Rint}
\end{center}
\end{figure}

Two regions within the Southern lobe of \cen\ were observed with the
\suz\ satellite \citep{Mitsuda07} from 2010 August 10 to 2010 August
12, as indicated in Figure\,\ref{fig:X-Rint} showing the 1.4\,GHz
total intensity map of the whole giant \cen\ structure \citep{Feain11}
with the two \suz\ pointings marked as blue squares. The observation
log, with details such as the nominal pointing positions and
observation times, is summarized in the first two rows of
Table\,\ref{tab:obslog} (``Lobe\,1'' and ``Lobe\,2'' exposures). In
this paper, we restrict the analysis to the data taken with the XIS
instrument \citep{Koyama07}, which consists of four CCD sensors each
placed on the focal plane of the X-ray Telescope
\citep[XRT;][]{Serlemitsos07}, and focus first
(\S\,\ref{sec:spots}--\ref{sec:spotmodel}) on faint X-ray features
likely associated, as argued below, with fine structure of the lobe
revealed by the most recent high-quality radio maps \citep{Feain11}.
Next (\S\,\ref{sec:diffuse}) we present the analysis of the diffuse
emission component filling the whole FOV of the instrument; for this
analysis we estimate the background from two shorter ``off''
\suz\ exposures, targeting the blank sky outside of (but close to) the
\cen\ Southern lobe, as summarized in the last two rows of
Table\,\ref{tab:obslog} (``Lobe\,3'' and ``Lobe\,4''; see also
Figure\,\ref{fig:rosat} below). At the time of the performed
observations the two front-illuminated CCDs (XIS0 and XIS3) and one
back-illuminated CCD (XIS1) were working well, so the data discussed
below for all the pointings were performed with all the XIS sensors
set to full-window and normal clocking modes.

In the following we describe the screening criteria for XIS data.
Events with \texttt{GRADE} 0, 2, 3, 4 and 6 were utilized in the data
reduction procedure, and flickering pixels were removed by using
\texttt{cleansis}. We selected good-time intervals by the same
criteria as described in the {\it The Suzaku Data Reduction
  Guide}\footnote{\texttt{http://heasarc.nasa.gov/docs/suzaku/analysis/abc/}},
\texttt{SAA\_HXD}$==$0 $\&\&$ \texttt{T\_SAA\_HXD}$>$436 $\&\&$
\texttt{ELV}$>$5 $\&\&$ \texttt{DYE\_ELV}$>$2 $\&\&$
\texttt{ANG\_DIST}$<$1.5 $\&\&$ \texttt{S0\_DTRATE}$<$3 $\&\&$
\texttt{AOCU\_HK\_CNT3\_NML\_P}$==$1. In the analysis we used the
\texttt{HEADAS} software version 6.11 and calibration database (CALDB)
released on 2012 February 11. The net exposures were 75.2\,ks and
81.4\,ks for the ``on'' pointings Lobe\,1 and Lobe\,2, respectively,
and about 20\,ks for the ``off'' pointings Lobe\,3 and Lobe\,4.

\begin{table*}[th]
{\footnotesize
\noindent
{\caption[] {\label{tab:obslog} Suzaku Observations of \cen\ in 2010}}
\begin{center}
\begin{tabular}{cccccc}
\hline\hline
\\
Target & Obs. ID & (R.A., Dec.) & Start time (UT) & End time (UT) & Net Exposure$^{\dagger}$\\
\\
\hline
\\
Lobe\,1 (``on'') & 705032010 & (200\deg.867, $-$45\deg.151) & Aug 10 15:52:16 & Aug 11 18:37:10 & 75.2 \\
Lobe\,2 (``on'') & 705033010 & (200\deg.394, $-$45\deg.136) & Aug 11 18:38:19 & Aug 12 23:08:19 & 81.4 \\
\\
Lobe\,3 (``off'') & 705034010 & (205\deg.750, $-$45\deg.132) & Aug 12 23:12:20 & Aug 13 08:17:12 & 20.1 \\
Lobe\,4 (``off'') & 705035010 & (196\deg.750, $-$45\deg.136) & Aug 13 08:21:01 & Aug 13 16:10:12 & 20.7 \\
\\
\hline\hline
\end{tabular}
\end{center}
$^{\dagger}$ After the event screenings, in ksec.
}
\end{table*}

\subsection{Compact X-ray Features}
\label{sec:spots}

\begin{figure*}[th]
\begin{center}
\includegraphics[width=\textwidth]{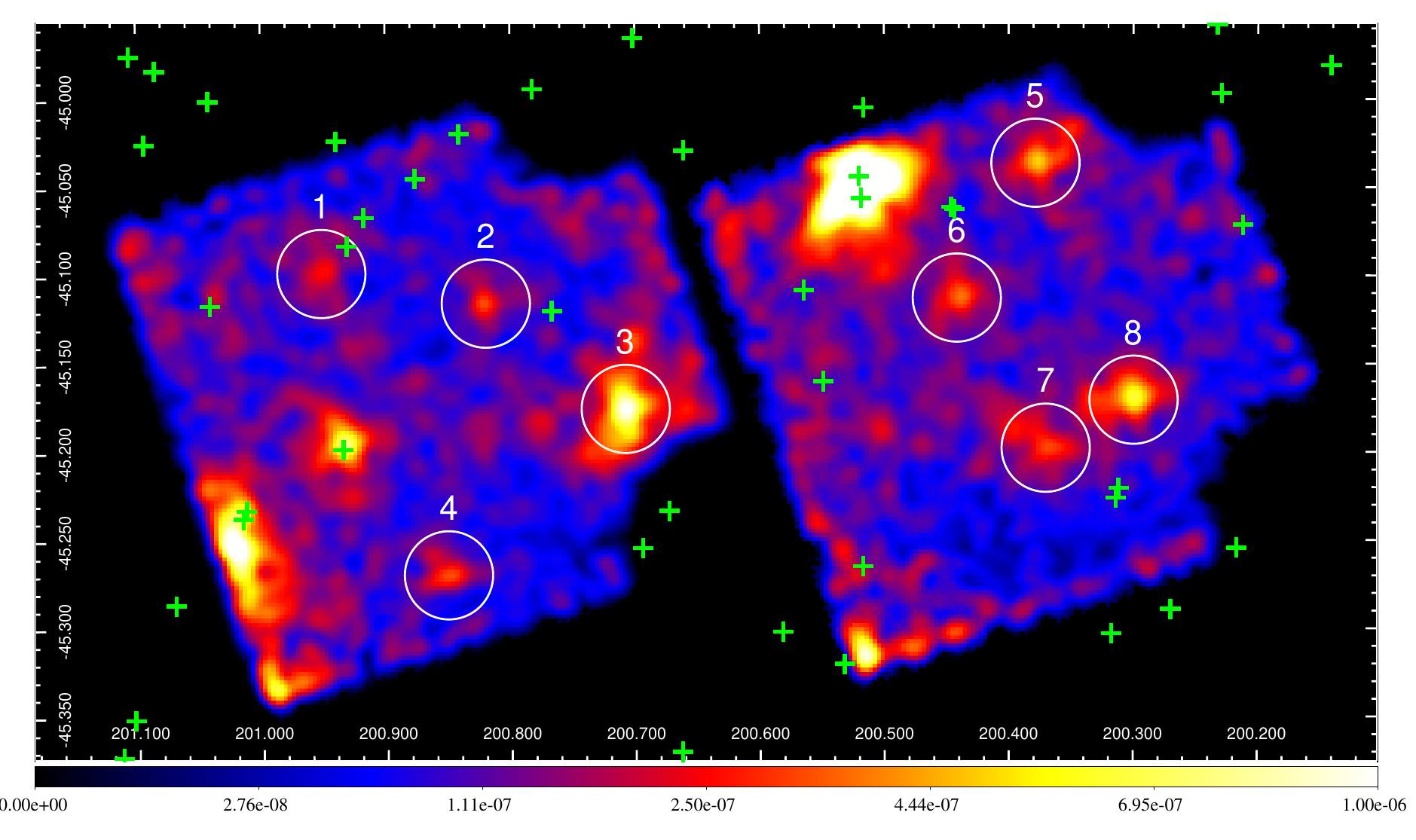}
\caption{\small \suz\ XIS0+3 image (after vignetting correction and subtraction of the NXB) of the targeted regions within the \cen\ Southern lobe corresponding to the photon energy range $0.5-10$\,keV. Green crosses mark the positions of the foreground/background sources from the SIMBAD database. White circles mark the flux extraction regions for the X-ray features which we consider as `possibly related to the lobe'. Spurious X-ray features are present in the bottom-left corners of the two FOVs, resulting from instrumental artifacts.}
\label{fig:suzaku}
\end{center}
\end{figure*}

Figure\,\ref{fig:suzaku} presents the \suz\ image of the targeted
``on'' regions within the \cen\ Southern lobe, corresponding to the
photon energy range $0.5-10$\,keV, with cataloged background or
foreground sources listed in the SIMBAD database marked by green
crosses. The figure corresponds to the combined XIS0 and XIS3 images
after the subtraction of the ``Non X-ray Background'' (NXB) and the
vignetting correction \citep[based on a flat image produced 
using \texttt{xissim};][]{Ishisaki07}. The maps were binned by a factor of eight 
in each $(x,\,y)$ direction and smoothed with a Gaussian kernel of four pixels. 
The NXB images were obtained from the night Earth data using \texttt{xisnxbgen} 
and corrected for the exposure using \texttt{xisexpmapgen} \citep{Tawa08}. 

As shown in the images, several bright X-ray spots are present within the FOV 
of the XIS instrument. The positions of three prominent spots closely match
the positions of point-like optical objects identified as stars:
HD\,116386 at (R.A.$=$201\deg.015, Dec.$=-$45\deg.2373), HD\,116335 at
(200\deg.9349, $-$45\deg.1980), and HD\,116099 at (200\deg.5186,
$-$45\deg.0564) together with nearby TYC\,8248-981-1 at (200\deg.5203,
$-$45\deg.0441); these three X-ray features are therefore considered
here as most likely unrelated to the lobe. We note that X-ray
counterparts for HD\,116386 and HD\,116099 are listed in the
\textit{ROSAT} All-Sky Survey (RASS) catalog as
1RXS\,J132403.1$-$451358 and 1RXS\,J132204.7$-$450312, respectively;
these, in fact, are the only cataloged X-ray sources within the area
covered by the two analyzed \suz\ pointings. The other X-ray features
detected with \suz\ at high significance level are missing any obvious
Galactic or extragalactic identifications (NED or SIMBAD databases),
and as such are considered below as `possibly related to the lobe';
these are marked in Figure\,\ref{fig:suzaku} with white circles and
denoted as ``src\,1--8''. It should be noted that spurious X-ray
features due to instrumental artifacts are present in the left-hand 
bottom corners of the two FOVs, and these are caused by the 
degradation of the CCD sensor XIS0 by the anomalous charge 
leakage\footnote{See the related discussion in the \suz\ 2010-01 
memo at \texttt{http://www.astro.isas.jaxa.jp/suzaku/doc/suzakumemo/}}.

For the spectral analysis of the eight X-ray features classified as
`possibly related to the lobe' we extracted the source fluxes from
circular regions with radii 1.5\am\ (well-matched to the PSF of the
XIS instrument) centered on each spot (see Figure\,\ref{fig:suzaku}),
and the background fluxes from larger ellipsoidal regions
``BGD/Lobe\,1'' and ``BGD/Lobe\,2'' chosen in the corresponding FOVs
to avoid any compact feature. The photon statistics for the
analyzed regions are given in Table\,\ref{tab:stat}. Only src\,3 and
5--8 are detected at a significance level above $5\sigma$.

With the given photon statistics it is impossible to state if the
`possibly related' spots are point-like for the XIS detector, or if
they are (marginally) extended. It is also not possible at this point
to state with high certainty if the detected X-ray features with no
obvious counterparts at optical wavelengths are indeed physically
associated with the giant \cen\ structure. A comparison of the radio
and X-ray morphologies presented in \S\,\ref{sec:spotmodel} below
indicates that this may indeed be the case, at least for some of them
(see also the discussion in \S\,\ref{sec:spot-origin}). And similarly,
one cannot exclude the possibility of a chance coincidence between an
X-ray spot associated with the giant lobe and a background/foreground
optical object. Because the exact positions of the detected X-ray
features cannot be measured given the angular resolution of the
XIS instrument and the photon statistics that we obtain, no meaningful
quantification of a chance probability for spurious association with
background/foreground object can be made. Here we consider an
association with such an object unlikely if all cataloged sources are
located more than 1.5\am\ away from the centers of the circular regions
used for the source flux extraction (see Table\,\ref{tab:stat}). For
the selected features 1--8 analyzed further below, this criterion is
marginally violated only in the case of src\,1, with the galaxy
2MASX\,J13234402$-$4505410 falling formally within the selected
\suz\ source region. The analysis presented in the section
\S\,\ref{sec:spotmodel} confirms, however, that the three prominent
X-ray spots coinciding with bright optical stars, and as such
classified as unrelated to the lobe, are spectrally distinct when
compared with the other detected features. These other features are
therefore not likely to be foreground stars, and instead we have to
consider the possibility that they are associated with uncataloged
background AGN. 

In order to evaluate the expected number of background AGN within the
FOVs of the two \suz\ ``on'' pointings, we note that the $\log N -
\log S$ distribution of such objects, constructed by \citet{Tozzi01}
based on the \textit{Chandra} Deep Field-South observations and
\textit{ROSAT} observations of the Lockman Hole, gives the number of
sources with $0.5-2$\,keV fluxes $>5 \times 10^{-14}$\,\Su\ between
20/deg$^{2}$ and 3/deg$^{2}$ (including the $1\sigma$ uncertainties
due to the sum of the Poisson noise). This implies a \emph{maximum} of
four background AGN within the two \emph{effective} FOVs of the XIS
instrument ($2 \times 0.09$/deg$^{2}$) at the accessible flux level, a
factor of at least two below the detected number of the X-ray spots
\emph{with no cataloged optical counterparts}. The number of
background X-ray sources evaluated by \citet{Tozzi01} in the hard
band, i.e. sources with $2-10$\,keV fluxes $>5 \times 10^{-14}$\,\Su,
is on the other hand higher, namely $\sim 40$/deg$^{2}$ \citep[as
  confirmed by other, more recent estimates, in agreement with the
  results of AGN population synthesis models;
  e.g.,][]{Ueda03,Gandhi03,Georgakakis08,Lehmer12}, meaning that we
would expect about seven AGN within the area of the giant lobe covered
by the \suz\ observations. This number is close to the number of our
detected X-ray spots with no obvious optical identification. Yet
unidentified background AGN are typically characterized by very flat
spectra around keV photon energies \citep[the average photon index
  $\Gamma = 1.4$;][]{Tozzi01}, while src\,1--8 discussed here display
on average much steeper X-ray continua (as we will show in
\S\,\ref{sec:spotmodel} below). These spots therefore do not have the
appearance of regular type 2 AGN lacking optical identification due to
a substantial intrinsic obscuration, with the possible exception of
src\,3. Moreover, we note that the two shorter ``off'' pointings with
\suz\ reveal no bright X-ray features at the flux level of src\,1--8
discussed here (which all lie well above the nominal limiting flux for
the 20\,ks exposure time of the `off' pointings), except for objects
with cataloged Galactic or extragalactic counterparts.

Hence, we conclude that while the contamination of the analyzed
\suz\ pointings with unidentified background AGN is an issue, there
seems to be a slight overabundance of relatively bright and
steep-spectrum X-ray sources in the ``on'' pointings when compared
both with the ``off'' pointings and also with the expected $\log N - \log S$
distribution of background AGN.

\begin{table}[!h]
{\footnotesize
\noindent
{\caption[] {\label{tab:stat} Photon Statistics for the Analyzed X-ray Spots}}
\begin{center}
\begin{tabular}{cccc}
\hline\hline
\\
Region & (R.A., Dec.)$^{\ast}$ & Total cnt. & Net cnt./\\
	&	&	&	stat.err.$^{\dagger}$	\\
\hline
\\
BGD/L1	&	---	&	1993/788$^{\star\star}$	&	--- \\
src\,1	&	(200\deg.949, $-$45\deg.097)	&	841	&	53/40 \\
src\,2	&	(200\deg.822, $-$45\deg.116)	&	890	&	102/41 \\
src\,3	&	(200\deg.708, $-$45\deg.177)	&	1627	&	839/49 \\
src\,4	&	(200\deg.851, $-$45\deg.270)	&	402$^{\ddag}$	&	4/28$^{\ddag}$ \\
\\
\hline
\\
BGD/L2	&	---	&	2370/781$^{\star\star}$	&	--- \\
src\,5	&	(200\deg.377, $-$45\deg.036)	&	1081	&	300/43 \\
src\,6	&	(200\deg.439, $-$45\deg.113)	&	1177	&	396/44 \\
src\,7	&	(200\deg.366, $-$45\deg.198)	&	1233	&	452/45 \\
src\,8	&	(200\deg.300, $-$45\deg.170)	&	1468	&	687/47 \\
\\
\hline\hline
\end{tabular}
\end{center}
$^{\ast}$ Center positions for the circular regions (radii $1.5$\am) used for the source flux extraction.\\
$^{\dagger}$ Statistical errors calculated assuming Poisson distribution as a square root of a sum of the source and background XIS0+3 counts, except for src\,4 ($^{\ddag}$) where only the XIS3 counts were considered (as the feature is located near the dead columns in XIS0).\\
$^{\star\star}$ Counts normalized to the effective area of the circular source regions.
}
\end{table}

In order to elaborate more on the possible association of the detected
X-ray spots with background AGN, we have investigated the available
mid-infrared (MIR) maps of the area covered by the \suz\ exposures
provided by the NASA Wide-field Infrared Survey Explorer satellite
\citep[\textit{WISE};][]{Wright10}. Over the course of a year after
its launch in late 2009, \wise\ carried out a deep all-sky survey in
four MIR bands, W1, W2, W3 and W4, centered on wavelengths around
3.4, 4.6, 12 and 22\,$\mu$m, respectively. The point-spread function
of the telescope corresponds to a Gaussian of about 6, 6, 6 and
12\as\ in W1--W4 respectively, sampled at 2.8, 2.8, 2.8 and 5.6\as/pix.
With nominal 5$\sigma$ point source sensitivies of $\sim$0.08, 0.1, 1
and 6\,mJy in the four bands, this survey is orders of magnitude more
sensitive than previous infrared all sky surveys
\citep{Soifer87,Ishihara10,Yamamura10}. The \wise\ pipeline implements
a sophisticated multi-band and multi-frame source detection algorithm.
The final product is an all-sky survey data release
catalog\footnote{\texttt{http://wise2.ipac.caltech.edu/docs/release/allsky}}
which reports Vega source magnitudes with the absolute calibration
being referenced to standard stars.

\begin{figure*}[!t]
\begin{center}
\includegraphics[width=5.5in]{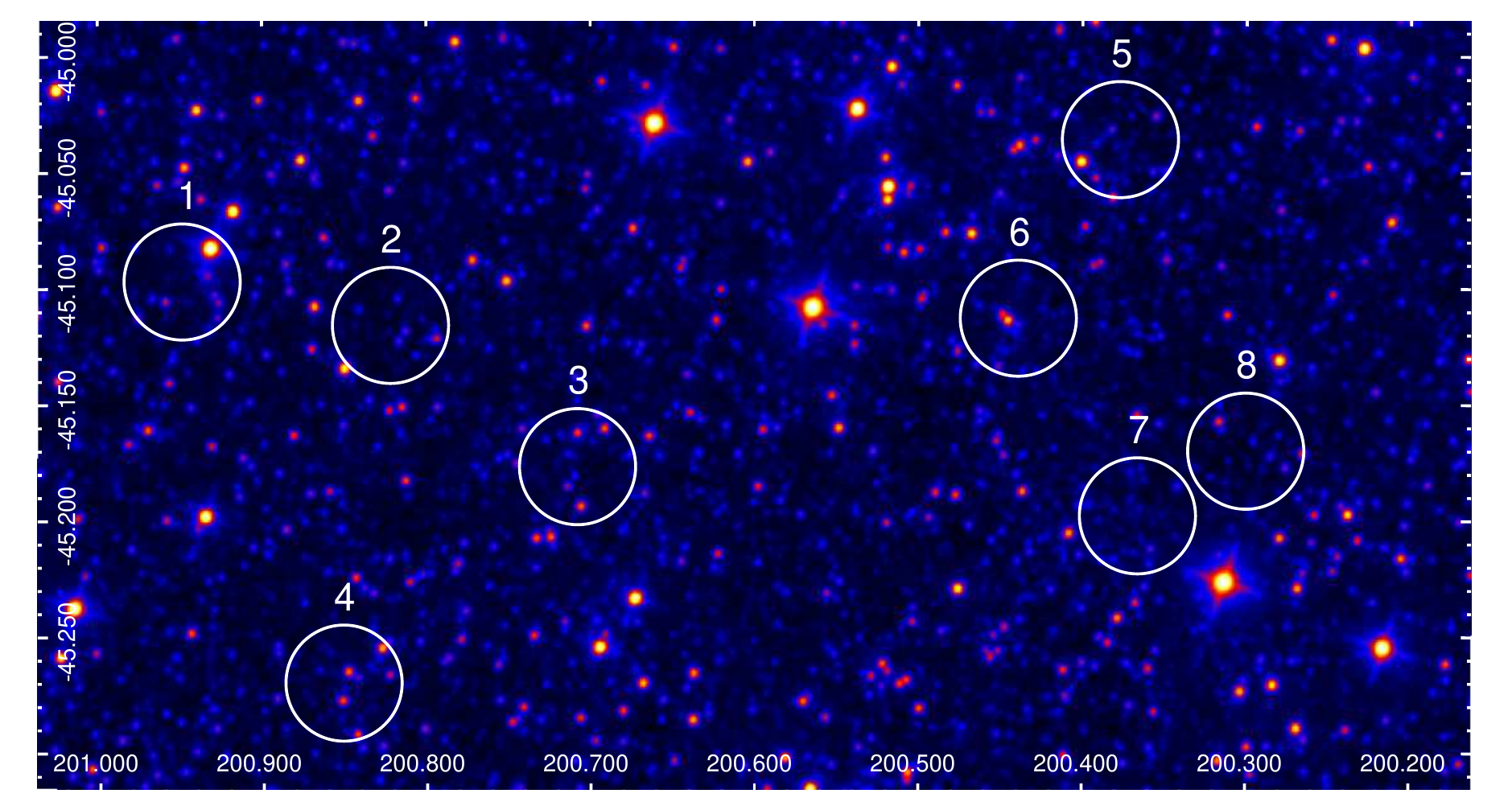}
\includegraphics[width=5.5in]{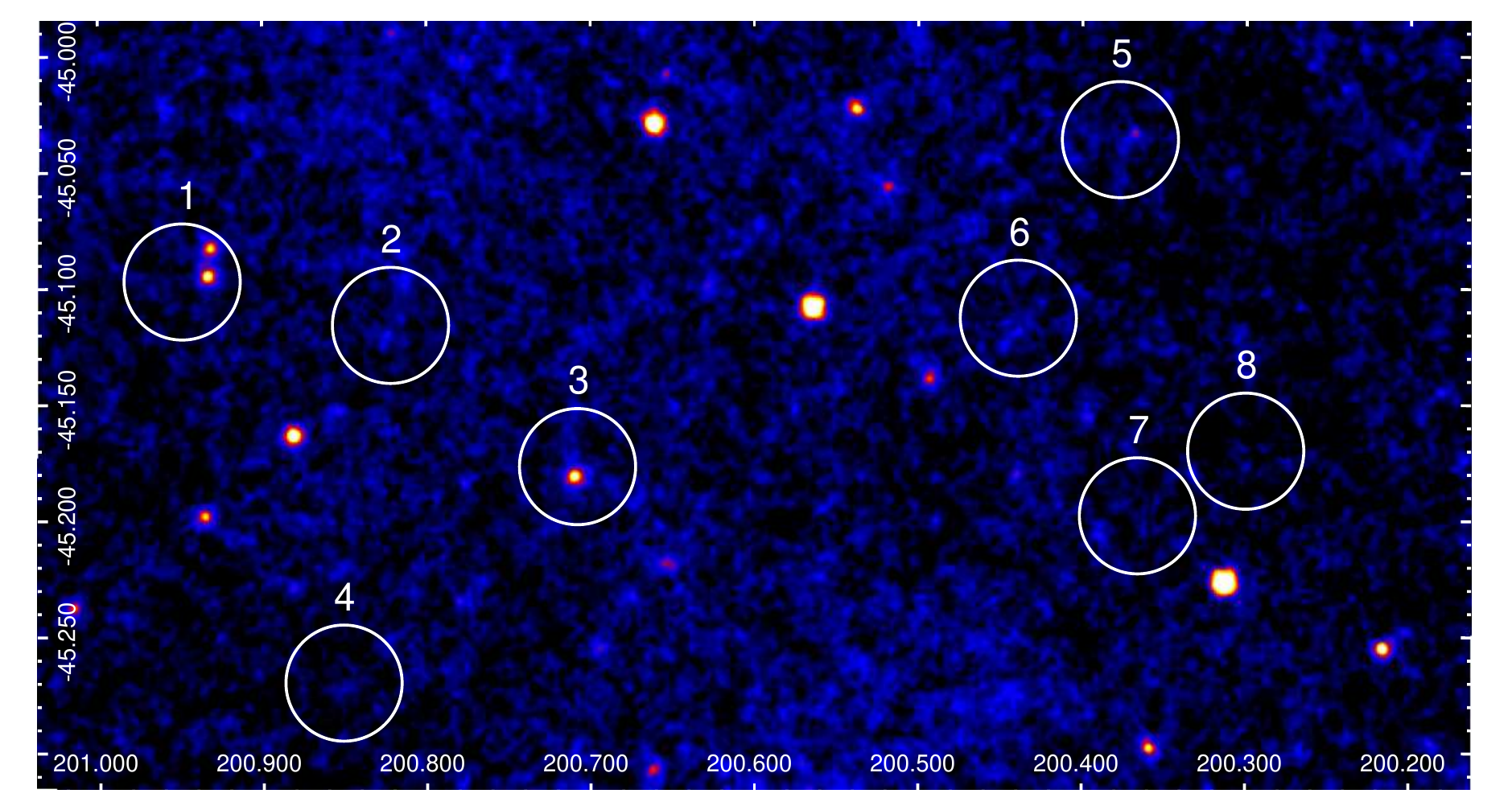}
\caption{\small \wise\ images of the part of the sky targeted by the two \suz\ ``on'' pointings at short (W1) and long (W4) mid-infrared wavelength (upper and lower panels, respectively). White circles denote the source extraction regions for the analyzed X-ray spots, as given in Table\,\ref{tab:stat} and shown in Figure\,\ref{fig:suzaku}.}
\label{fig:wise}
\end{center}
\end{figure*}

The corresponding images in the short (W1) and long (W4) wavelength
bands are shown in Figure\,\ref{fig:wise}. The all-sky data release
was searched for detected objects within a radius of 1.5\am\ around
the positions of the region centers listed in Table\,\ref{tab:stat}.
All sources with the photometric quality (\texttt{ph\_qual}) flag of
A, B or C, denoting real (albeit including low signal--to--noise)
detections were included. The selection was carried out for each band
individually. We used profile-fit magnitudes and standard zeropoints
\citep{Jarrett11} in order to measure the approximate contribution due
to associated point sources. The summed fluxes of detections for each
region are listed in Table\,\ref{tab:wise}. These should be treated as
upper limits on the contribution of point sources at the position of each X-ray feature discussed here.

Figure\,\ref{fig:wise} shows that the areal density of
\wise\ detections in the targeted part of the sky decreases strongly
with wavelength, with each source extraction region for all the
analyzed X-ray spots overlapping with several MIR emitters at
$3.4$\,$\mu$m, but being typically devoid of any bright sources at
$22$\,$\mu$m. Only the regions of src\,1 and 3 are completely
dominated by a couple of sources alone. In particular, src\,1
coincides with one very blue object at (R.A.$=$200\deg.9314,
Dec.$=-$45\deg.0827), a star, TYC\,8248-535-1, present in the
SIMBAD database, and one red object at
(200\deg.9335, $-$45\deg.0947), the aforementioned galaxy 2MASX
J13234402$-$4505410. The brightest MIR emitter in the region of src\,3 is a
very red source located at (200\deg.7099, $-$45\deg.1809), lacking any
optical identification.

The last two columns in Table\,\ref{tab:wise} list the corresponding
MIR--to--X-ray spectral indices, computed formally with the given
$3.4$\,$\mu$m and $22$\,$\mu$m fluxes, and $1$\,keV fluxes of each
spot (see \S\,\ref{sec:spotmodel} below). The provided numbers do not
correspond necessarily to real source spectra, as the utilized
\wise\ fluxes are superpositions of contributions from distinct
objects, especially at shorter wavelengths. We note however that the
evaluated $\alpha_{\rm 3.4\,\mu m-1\,keV}$ indices are relatively
steep, $\gtrsim 1.7$, so even assuming that the W1-band fluxes of real
counterparts of the X-ray features are one order of magnitude lower
than the listed values, these indices would still all be $> 1$. The
situation is somewhat different when $\alpha_{\rm 22\,\mu m-1\,keV}$ are
considered, since in the W4 band most of the analyzed regions
(including X-ray bright src\,6--8) are lacking significant detections.
These non-detections may potentially challenge the possibility for AGN
associations, as the MIR--to--X-ray power-law slopes for local
radio-quiet AGN have been shown to be $\gtrsim 1$
\citep[e.g.,][]{Gandhi09}.

\begin{table*}[!th]
{\footnotesize
\noindent
{\caption[] {\label{tab:wise} Total MIR Fluxes (in milli-Jy) for the Source Extraction Regions of the X-ray Spots}}
\begin{center}
\begin{tabular}{lrrrrrr}
\hline\hline
\\
Region & W1[3.4\,$\mu$m] & W2 [4.6\,$\mu$m] & W3 [12\,$\mu$m] & W4$^{\star}$ [22\,$\mu$m] & $\alpha_{\rm 3.4\,\mu m-1\,keV}$ & $\alpha_{\rm 22\,\mu m-1\,keV}$\\
\\
\hline
\\
src\,1$^{\ast}$	&	287.8 & 141.0 & 34.2 & 17.7 & 2.2 & 1.5 \\
src\,2 & 10.9 & 5.9 & 3.7 &  $<$6 & 1.8 & $<$1.4\\
src\,3$^{\dagger}$ & 33.7 & 18.1 & 10.7 & 14.8 & 1.8 & 1.4\\
src\,4 & 49.1 & 25.3 & 4.8 & $<$6 & 2.0 & $<$1.4\\
src\,5 & 58.6 & 30.2 & 7.1 & 8.6 & 2.0 & 1.4\\
src\,6 & 35.1 & 17.8 & 7.0 & $<$6 & 1.9 & $<$1.4\\
src\,7 & 11.4 & 6.2 & 3.1 & $<$6 & 1.7 & $<$1.3\\
src\,8 & 17.6 & 9.4 & 2.1 & $<$6 & 1.8 & $<$1.3\\
\\
\hline\hline
\end{tabular}
\end{center}
$^{\star}$ For the W4 filter, several regions are devoid of detections; in these cases, the nominal 5$\sigma$ point source detection limit of 6\,mJy has been assumed.\\
$^{\ast}$ The total \wise\ flux is dominated at short wavelengths by a star TYC\,8248-535-1, and at longer wavelengths by a galaxy 2MASX J13234402$-$4505410.\\
$^{\dagger}$ The total \wise\ flux is dominated by a single red object located at (200\deg.7099, $-$45\deg.1809) lacking any optical identification.
}
\end{table*}

All in all, based on the analysis of the \wise\ data we conclude that
at least src\,3 --- clearly coinciding with a red MIR emitter --- is
consistent with being a background obscured AGN rather than a feature
related to the lobe. This conclusion is supported by the X-ray
spectral analysis presented in the next section below. In addition,
the relatively dim src\,1 could be considered as a likely X-ray
counterpart of a `Low-Luminosity AGN'. As 
for the other X-ray spots considered, the situation is still open, especially as
more AGN associations could be expected on statistical grounds. We
draw attention to the particularly interesting cases of src\,7 and 8,
which lack any optical or MIR counterparts, but at the same time, as
shown below, are characterized by relatively steep X-ray spectra and
positionally associated with a prominent radio filament in the lobe.

\subsection{X-ray Spots: Spectral Analysis}
\label{sec:spotmodel}

\begin{table*}[th]
{\footnotesize
\noindent
{\caption[] {\label{tab:fits} Modeling Results for the X-ray Spots Detected at $>5 \sigma$ Level}}
\begin{center}
\begin{tabular}{cccccccccccc}
\hline\hline
\\
source & red.$\chi^2$/dof & $\Gamma$ & $F_{\rm x,\,abs}$ & $F_{\rm x}$ & red.$\chi^2$/dof & $kT$ & $F_{\rm x,\,abs}$ & $F_{\rm x}$ & $Z/Z_{\odot}$ & Norm & red.$\chi^2$/dof \\
(1) & (2) & (3) & (4) & (5) & (6) & (7) & (8) & (9) & (10) & (11) & (12) \\
\\
\hline
\\
src\,3 & 1.340/8 & $1.32^{+0.12}_{-0.12}$ & 29.2 & 30.4 & 0.671/51 & 32.8 & 28.3 & 29.5 & 0.3$^f$ & $17.1^{+3.6}_{-2.2}$ & 0.639/52\\
src\,5 & 0.211/8 & $2.00^{+0.35}_{-0.32}$ & 6.6 & 7.3 & 1.192/30 & $4.6^{+4.7}_{-1.6}$ & 6.3 & 6.8 & 0.3$^f$ & $4.77^{+0.81}_{-0.82}$ & 1.249/31\\
src\,6 & 0.395/8 & $1.89^{+0.24}_{-0.21}$ & 7.5 & 8.3 & 0.804/36 & $4.6^{+2.6}_{-1.3}$ & 6.9 & 7.4 & 0.3$^f$ & $5.24^{+0.66}_{-0.67}$ & 0.788/37\\
src\,7 & 0.693/8 & $2.52^{+0.24}_{-0.23}$ & 7.1 & 8.3 & 1.020/38 & $2.1^{+0.7}_{-0.3}$ & 5.8 & 6.5 & 0.3$^f$ & $6.55^{+0.69}_{-0.69}$ & 1.373/39\\
src\,8 & 0.598/8 & $1.68^{+0.14}_{-0.14}$ & 15.8 & 16.8 & 0.685/47 & $7.3^{+4.3}_{-2.1}$ & 14.9 & 15.8 & 0.3$^f$ & $9.57^{+0.73}_{-0.71}$ & 0.770/48\\
\\
\hline\hline
\end{tabular}
\end{center}
(1) source ID; (2) quality of a constant fit to the source lightcurve
(reduced $\chi^2$ value/degree of freedom); (3) photon index in the
\texttt{PL} model with $90\%$ confidence level errors; (4) absorbed
$0.5-10$\,keV flux of the source in the \texttt{PL} model, in
units of $10^{-14}$\,\Su; (5) unabsorbed $0.5-10$\,keV flux of the
source in the \texttt{PL} model in the same units; (6) quality of the
\texttt{PL} fit; (7) plasma temperature for the \texttt{APEC} model in
units of keV with $90\%$ confidence level errors (except for src\,3 for which no meaningful errors could be calculated); (8) absorbed $0.5-10$\,keV flux of the source in the \texttt{APEC} model, in units of $10^{-14}$\,\Su; (9) unabsorbed $0.5-10$\,keV flux of the source in the \texttt{APEC} model in the same units; (10) fixed (``$f$'') abundance in the \texttt{APEC} fit; (11) \texttt{APEC} normalization $\times 10^{-5}$; (12) quality of the \texttt{APEC} fit.
}
\end{table*}

\begin{figure}[!h]
\begin{center}
\includegraphics[width=2.15in]{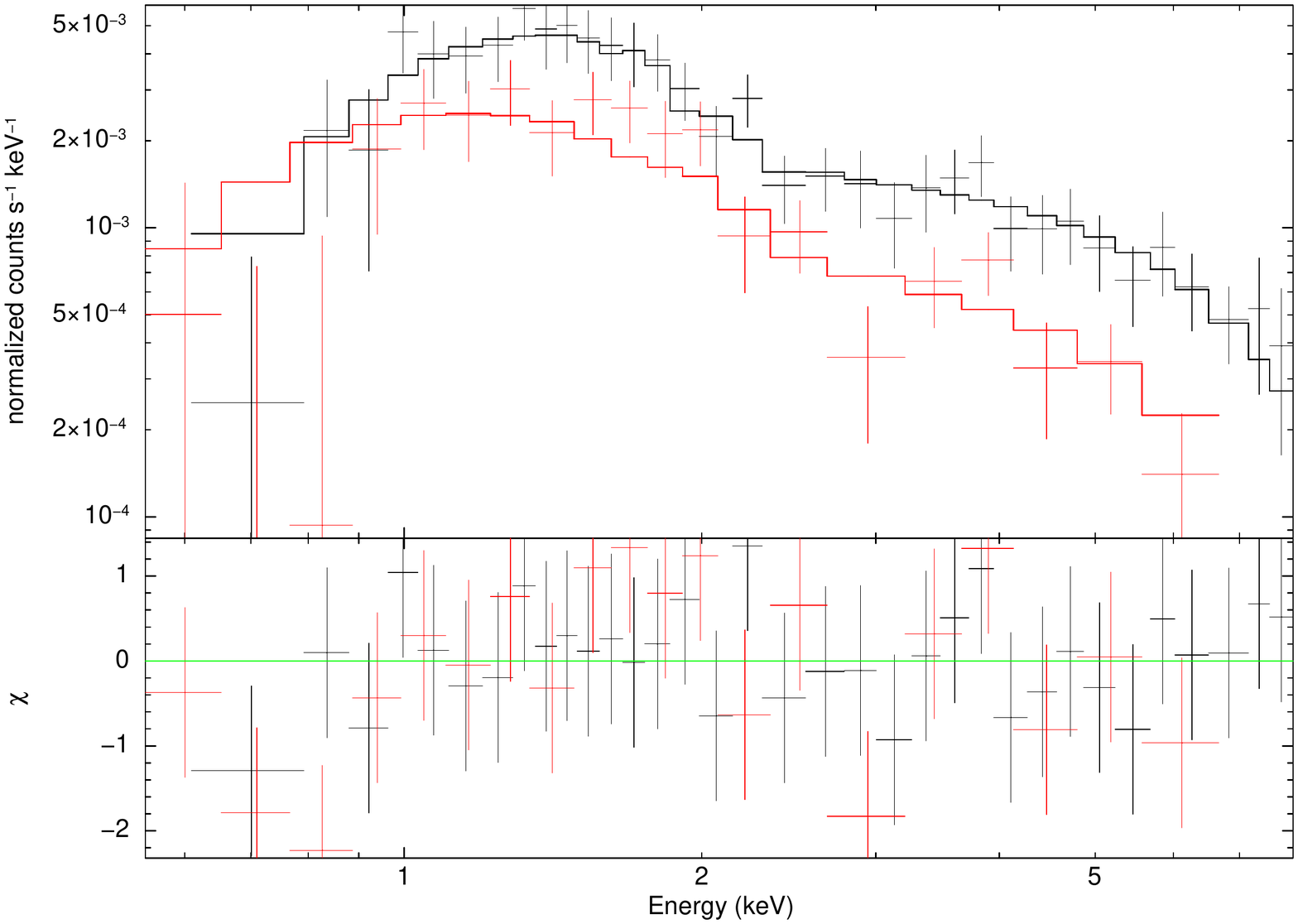}
\includegraphics[width=2.15in]{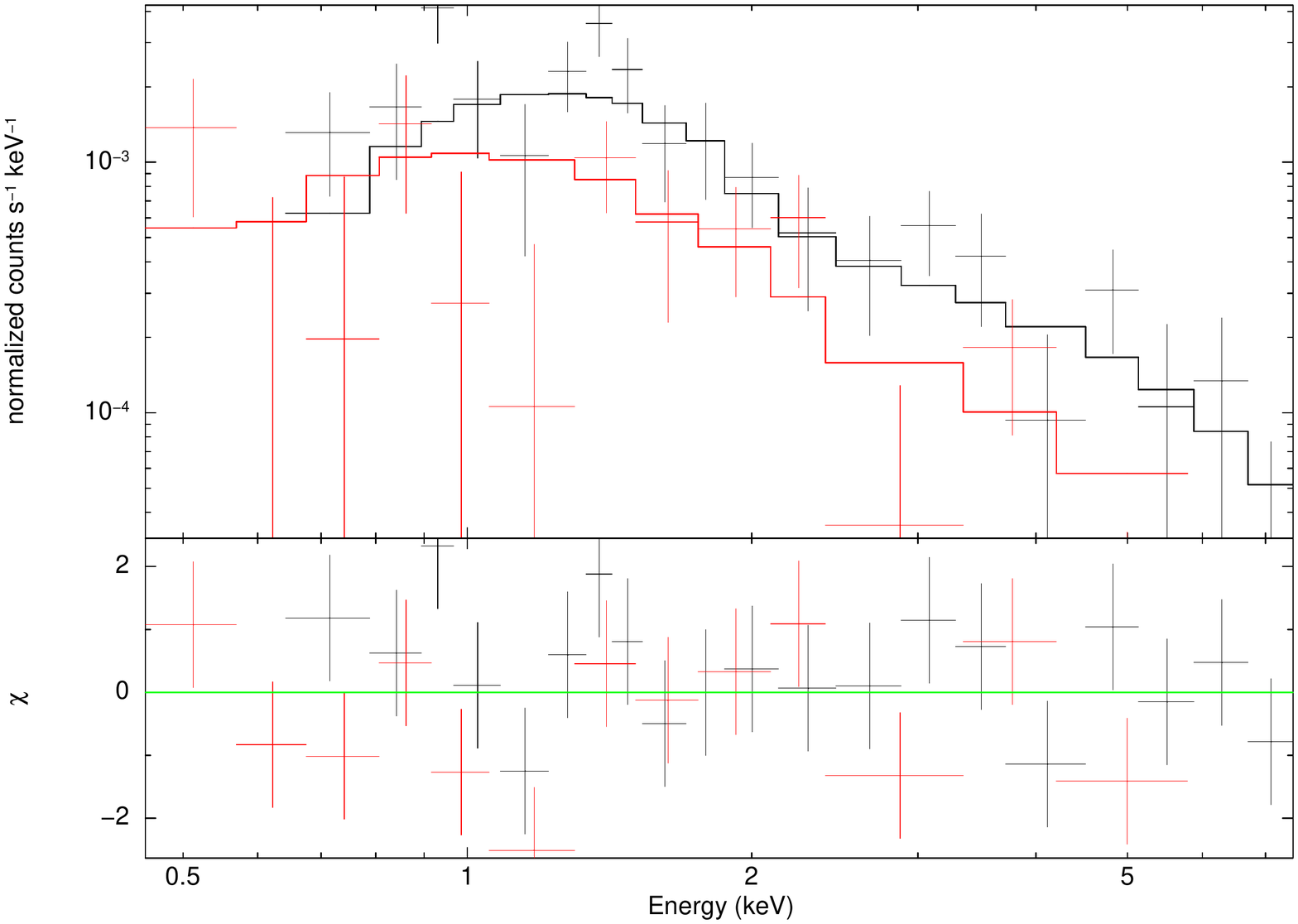}
\includegraphics[width=2.15in]{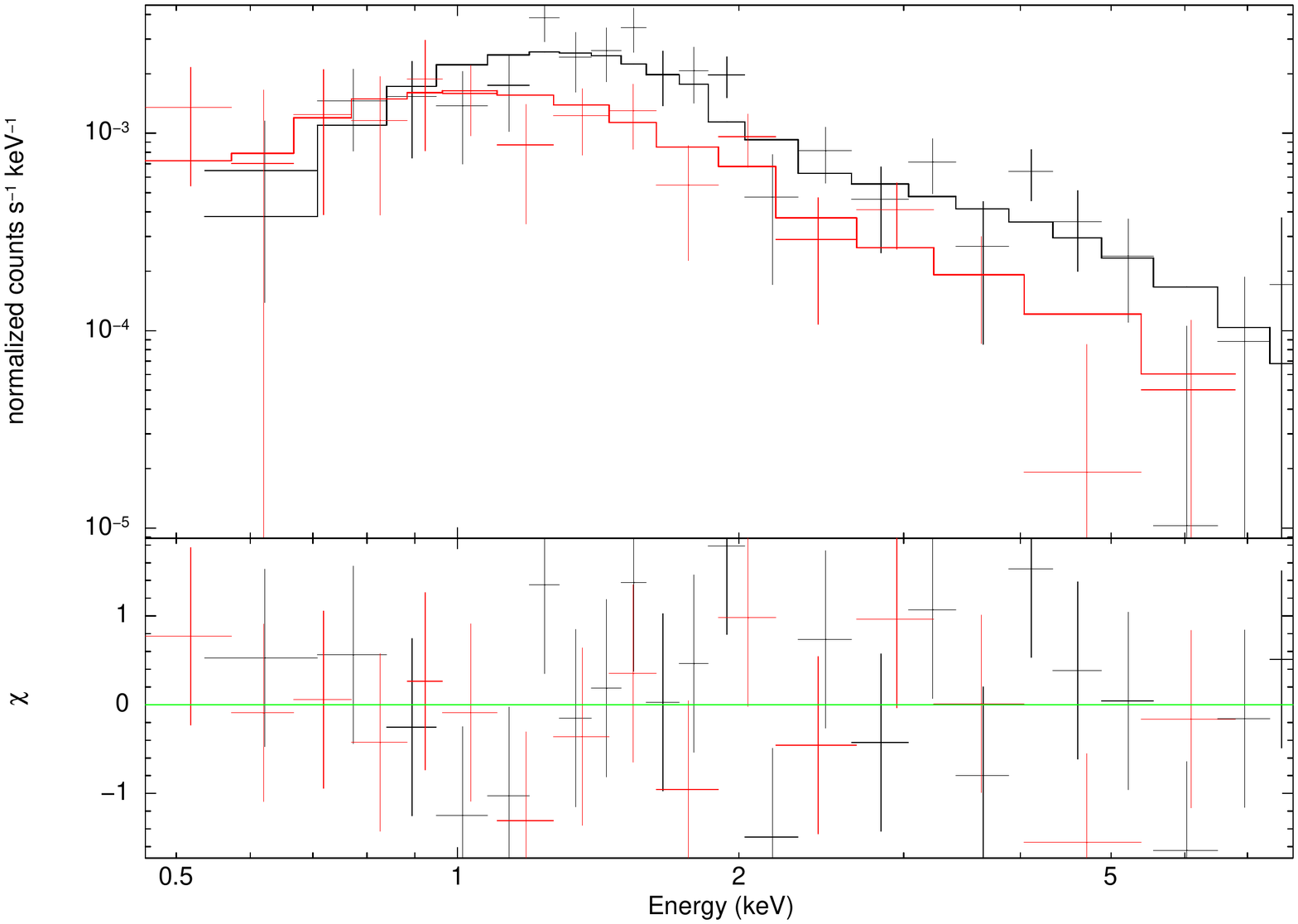}
\includegraphics[width=2.15in]{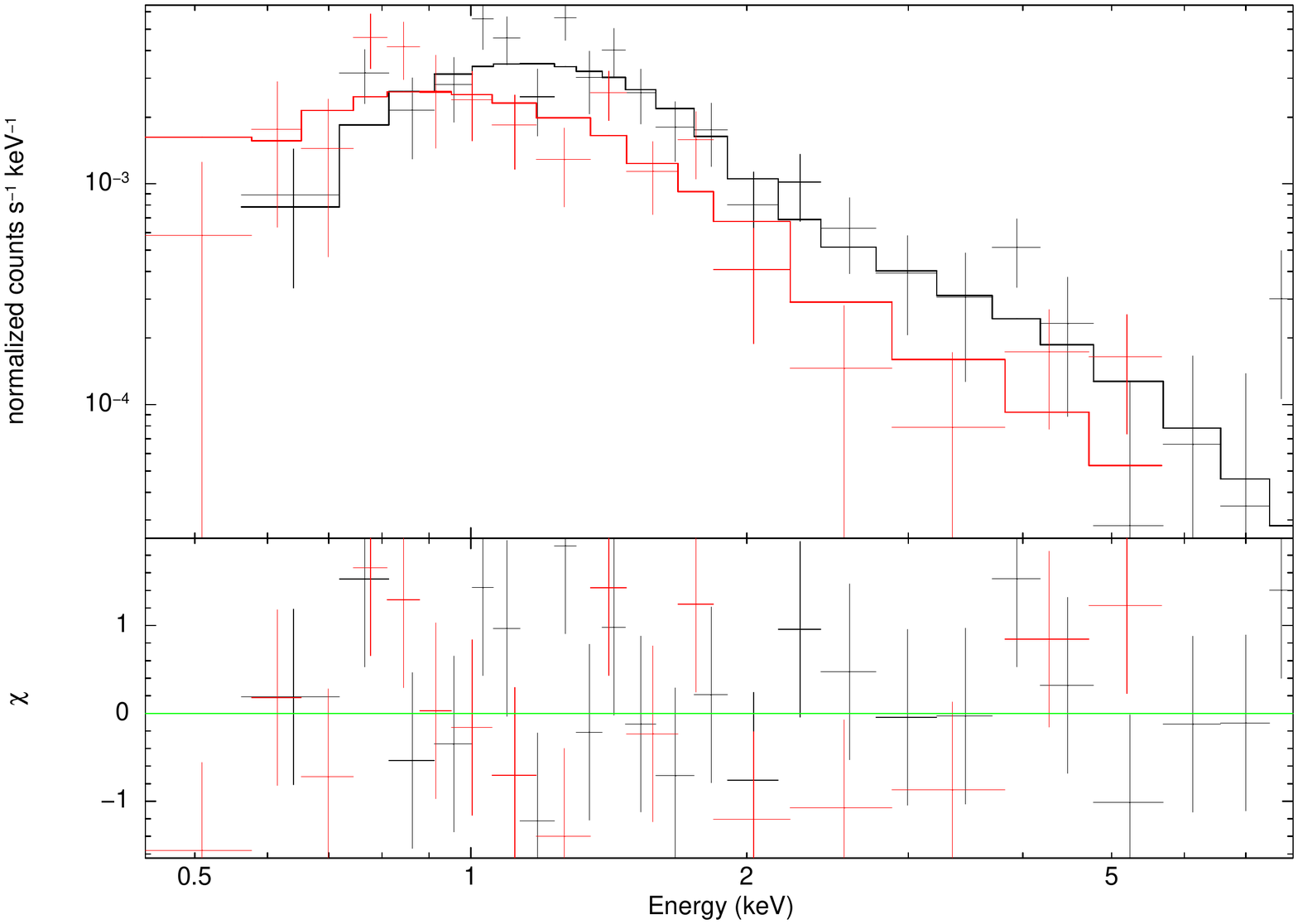}
\includegraphics[width=2.15in]{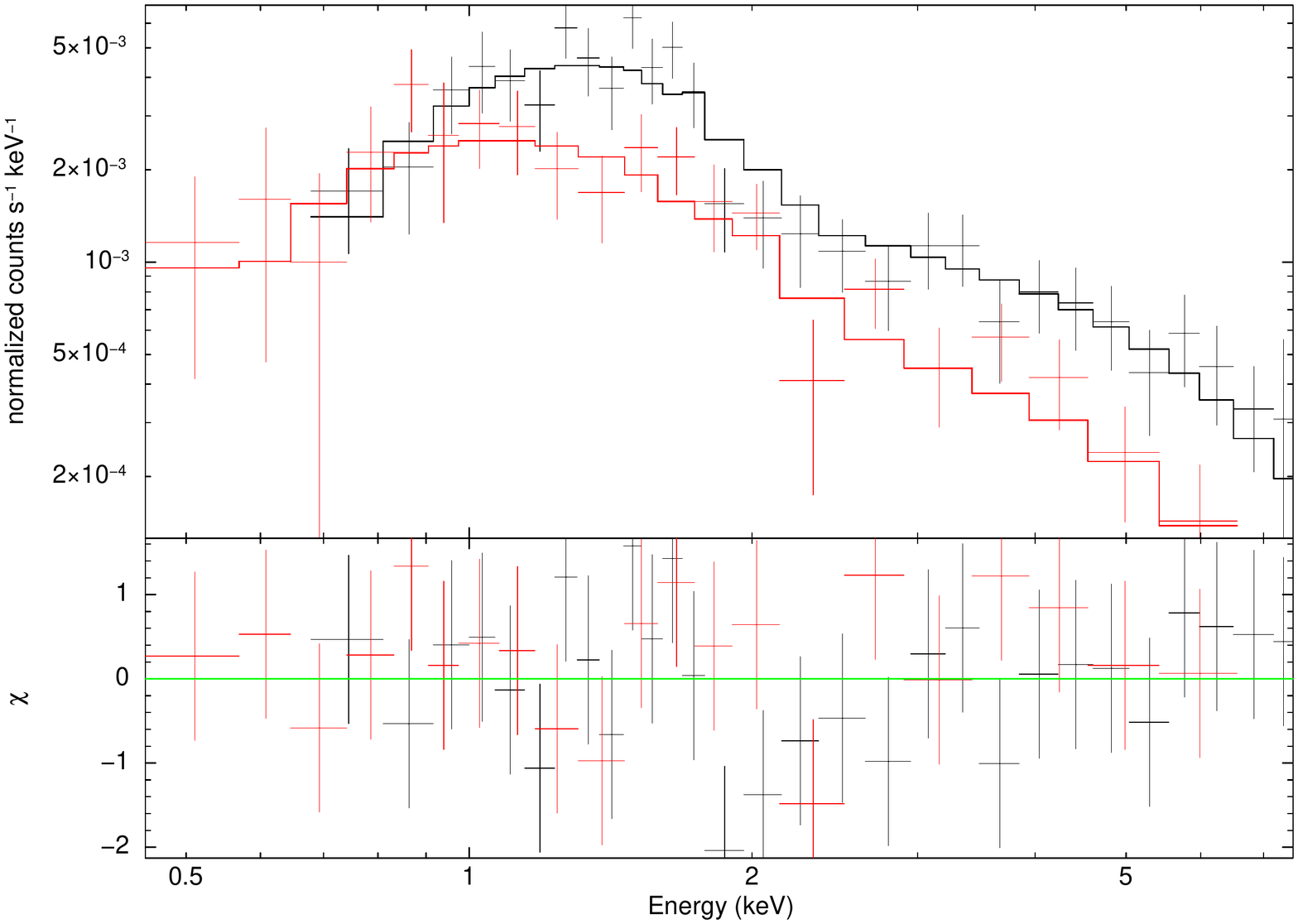}
\caption{\small X-ray spectra of sources 3, 5--8 (from top to bottom, respectively), as measured by front-illuminated CCDs XIS0+3 (black datapoints) and back-illuminated CCD XIS1 (red datapoints). Continua represent \texttt{PL} models fitted to the data (Table\,\ref{tab:fits} and \S\,\ref{sec:spotmodel}). The datapoints are binned in 40\,cts.}
\label{fig:Xspec}
\end{center}
\end{figure}

\begin{figure}[!h]
\begin{center}
\includegraphics[width=2.16in]{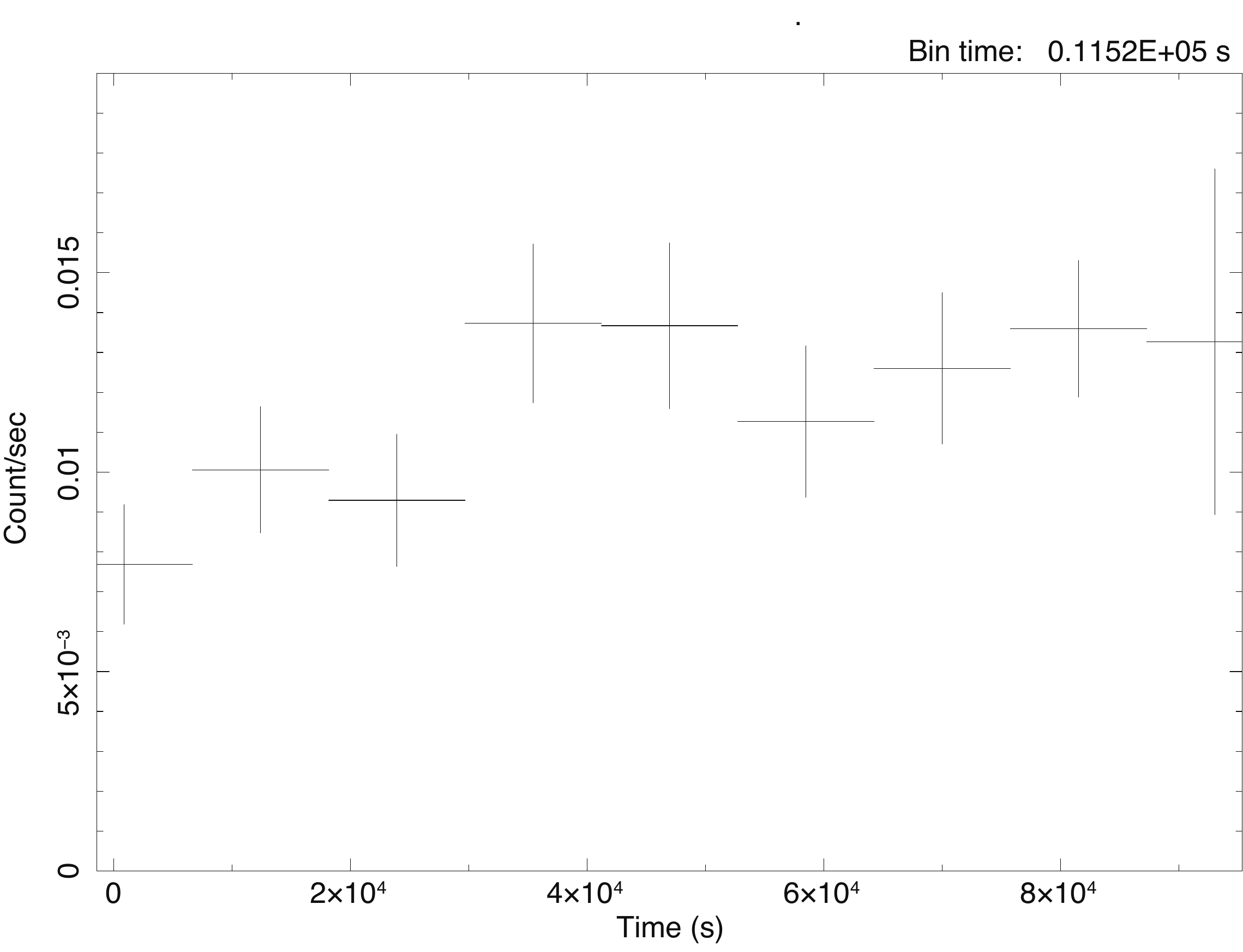}
\includegraphics[width=2.16in]{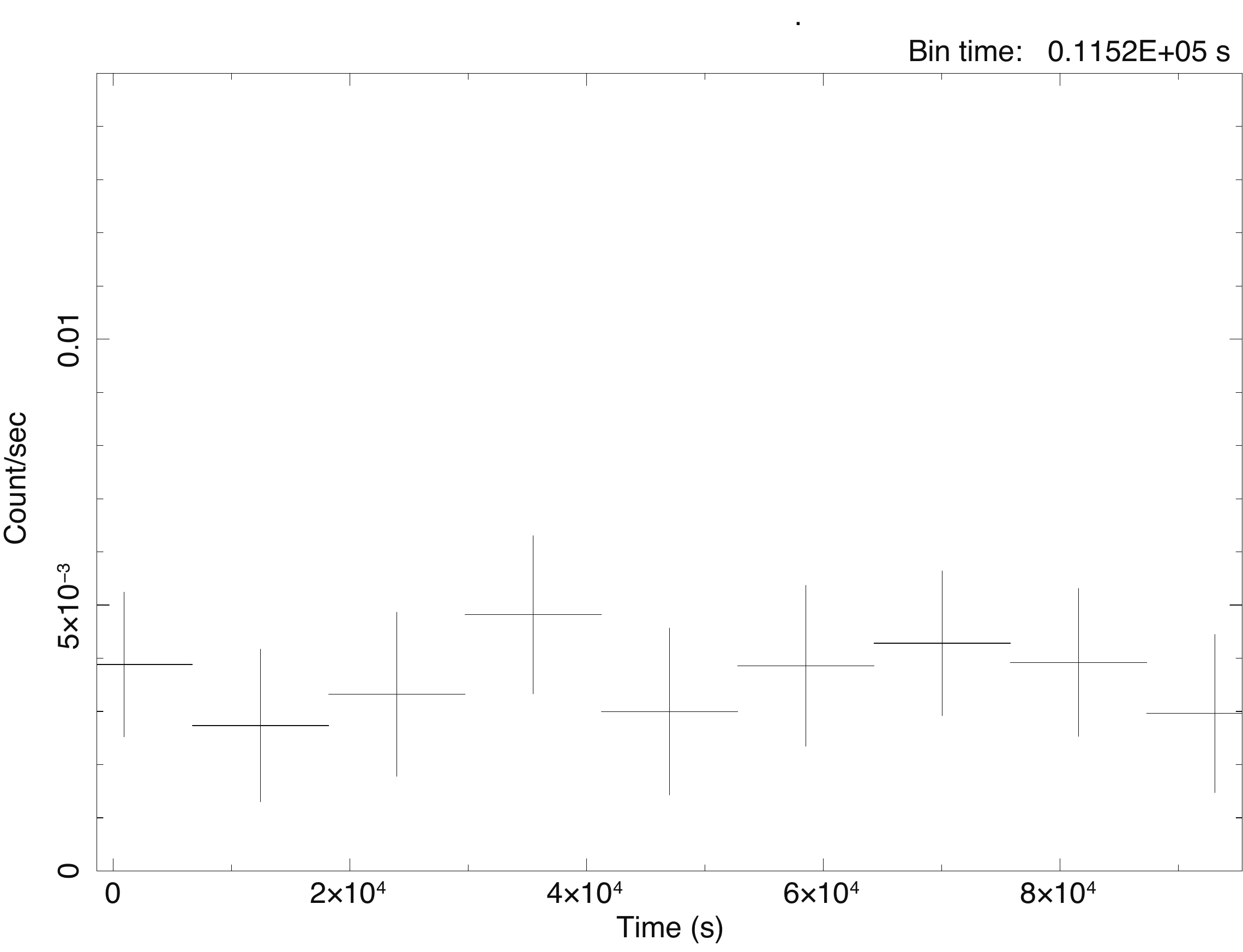}
\includegraphics[width=2.16in]{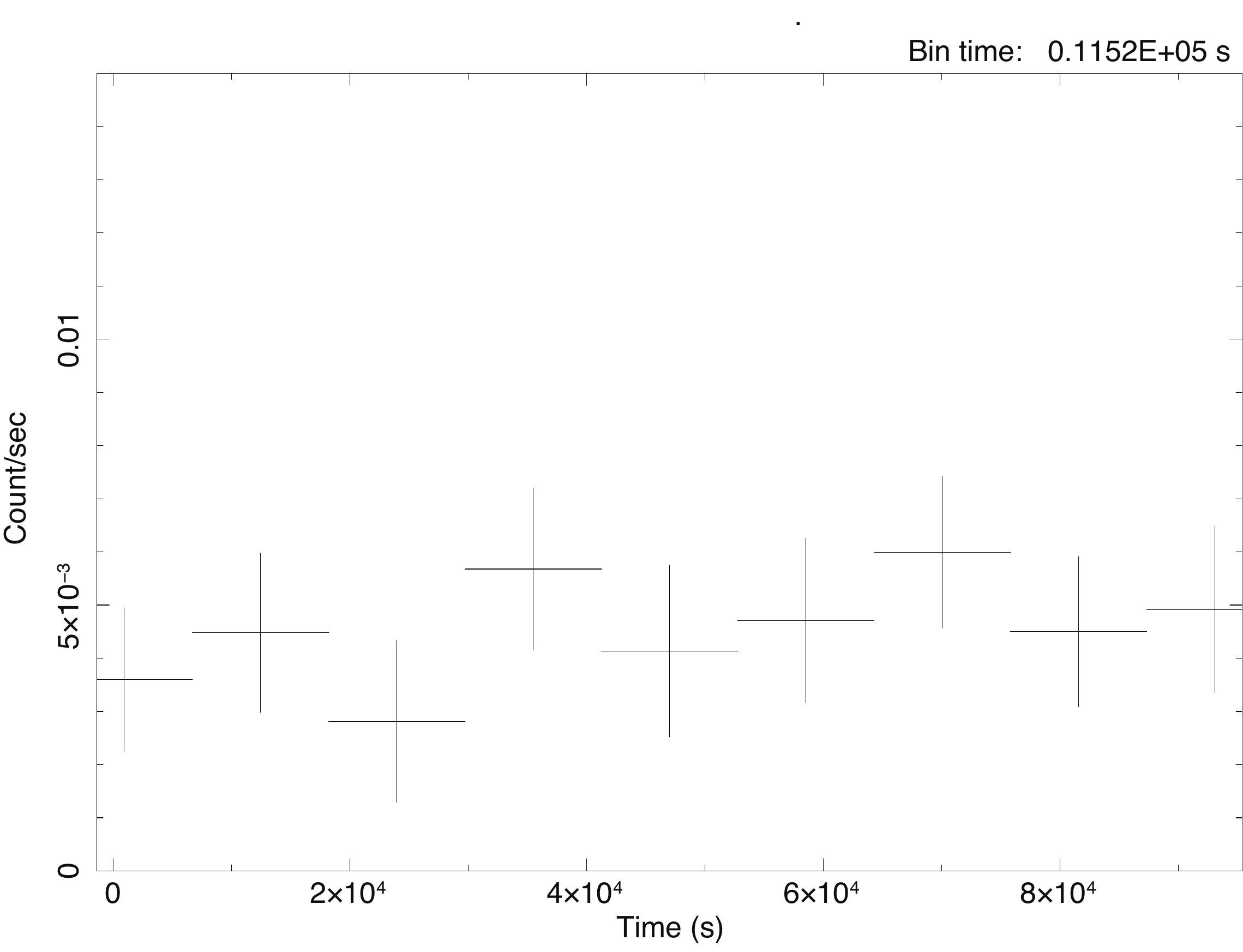}
\includegraphics[width=2.16in]{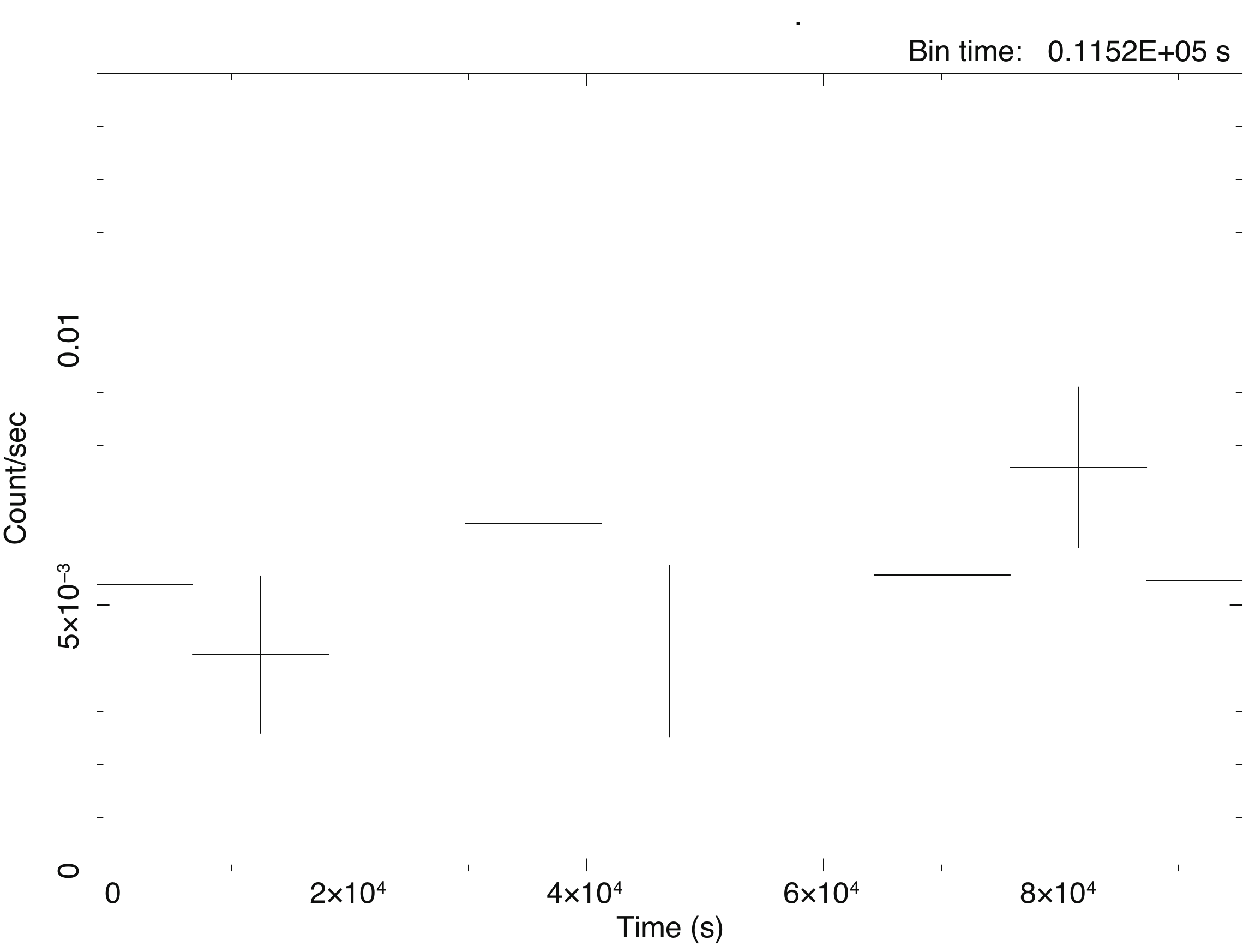}
\includegraphics[width=2.16in]{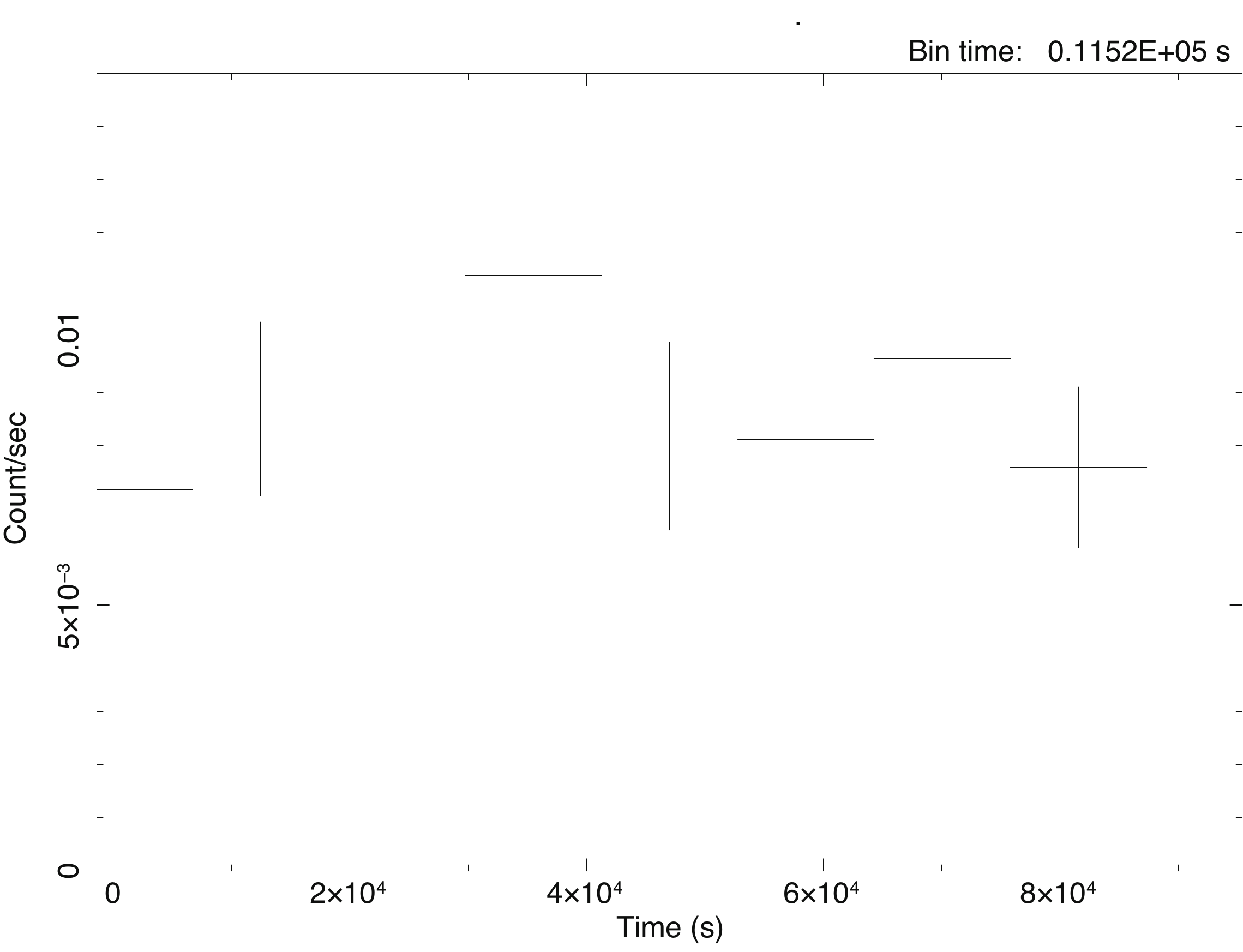}
\caption{\small X-ray lightcurves of sources 3, 5--8 (from top to bottom, respectively) during the corresponding \suz\ exposures, binned in 3.2\,h intervals. The performed constant fits to the lightcurves imply steady emission during the performed exposures (see Table\,\ref{tab:fits} and \S\,\ref{sec:spotmodel}).}
\label{fig:Xvar}
\end{center}
\end{figure}

We fitted the spectra of the X-ray features detected at significance
levels above $5\sigma$ (src\,3 and 5--8) using \texttt{XSPEC}
(v12.7.0) with an absorbed power-law model (\texttt{PL}) and an
absorbed thermal model \texttt{APEC} \citep{APEC}, in all cases
freezing the absorbing neutral hydrogen column density at the Galactic
value in the direction of the source, $N_{\rm H,\,Gal} = 0.7 \times
10^{21}$\,cm$^{-2}$. The results of the fitting, summarized in
Table\,\ref{tab:fits}, indicate that both models provide an equally
good description of the data, with the emerging photon indices $\Gamma
\simeq 2.0 \pm 0.5$ for all the spots, except for src\,3 which is
characterized by $\Gamma <1.5$, or equivalently with plasma
temperatures within the range $kT \sim 1-10$\,keV, again except for
src\,3 which has a poorly constrained $kT > 10$\,keV.
Figure\,\ref{fig:Xspec} presents the \suz\ spectra of src\,3 and 5--8
and the corresponding \texttt{PL} fits. The summed unabsorbed
$0.5-10$\,keV flux of all the analyzed features is $F_{\rm X} \simeq 7
\times 10^{-13}$\,\Su, implying a total $0.5-10$\,keV luminosity of
the spots $L_{\rm X} \sim 10^{39}$\,\Lu\ (for the assumed distance of
$D = 3.7$\,Mpc). These values correspond to a surface area $2 \times
0.09$\,deg$^2$, i.e. twice the \emph{effective} FOV of the XIS
instrument. All the analyzed X-ray sources appear steady during the
performed exposures, as shown in Figure\,\ref{fig:Xvar} and quantified
in Table\,\ref{tab:fits}. We note that in the fitting procedure using
the \texttt{APEC} model, due to the rather limited photon statistics
the plasma abundance could not be treated as a free parameter, but
instead was fixed at the value $Z = 0.3\, Z_{\odot}$ \citep[as roughly
  expected for the inter-galactic medium at larger distances from the
  galaxy group/cluster center; see, e.g.,][]{Sun12}. Alternative fits
with the abundance reduced to $Z = 0.1 \, Z_{\odot}$, or even assuming
a pure bremsstrahlung model instead of \texttt{APEC}, returned essentially the same model parameters (including normalization), and comparable $\chi^2$ values.

As follows from Table\,\ref{tab:fits}, src\,3 stands out in terms of its flux and spectral properties from the other X-ray spots analyzed here, displaying a flat X-ray spectrum with no signatures for curvature or a cut-off up to 10\,keV photon energies. This justifies further the idea, presented previously in \S\,\ref{sec:spots} based on the analysis of the \wise\ data, that this feature is related to a background type 2 AGN rather than to the \cen\ giant lobe\footnote{We note some hints for the presence of a line-like feature/excess in the \suz\ spectrum of src\,3 (both front- and back-illuminated CCDs), around the observed photon energies of $\simeq 4$\,keV. If due to iron K$\alpha$ emission, this feature would be consistent with a source redshift of, roughly, $z \sim 0.5$.}. With src\,3 excluded, the summed $0.5-10$\,keV fluxes and luminosities of the modeled src\,5--8 are $F_{\rm X} \simeq 4 \times 10^{-13}$\,\Su\ and $L_{\rm X} \sim 0.6 \times 10^{39}$\,\Lu, respectively, meaning $F_{\rm x} \sim 10^{-13}$\,\Su\ and $L_{\rm x} \sim 10^{38}$\,\Lu\ per spot, on average.

\begin{figure*}[th]
\begin{center}
\includegraphics[width=7.0in]{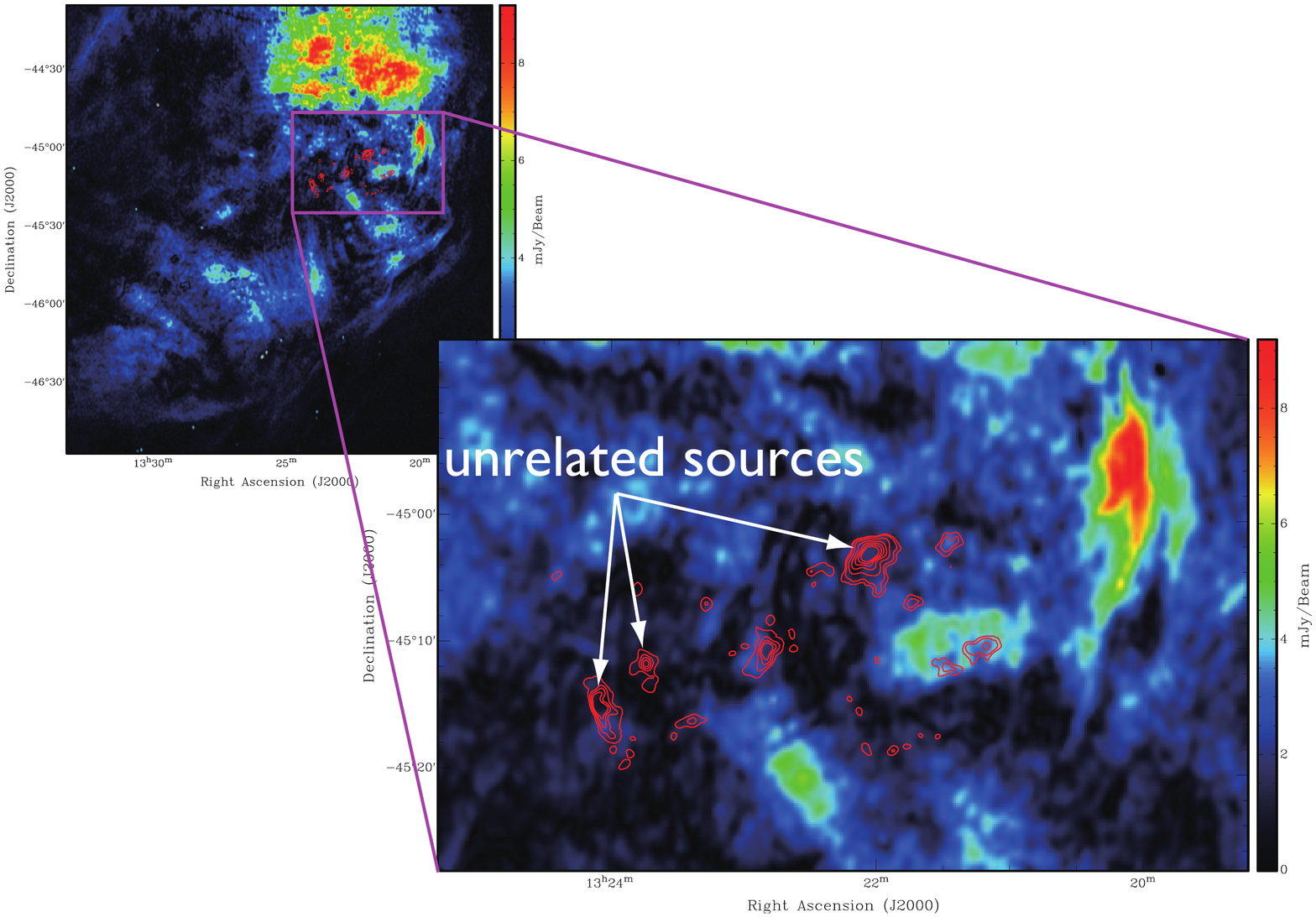}
\caption{\small 
Polarized radio intensity map of the \cen\ Southern lobe with the X-ray \suz\ countours superimposed. Note that the most pronounced radio filament in the top-right corner of the zoomed image was unfortunately not covered by the \suz\ observations.
}{\label{fig:X-Rpol}}
\end{center}
\end{figure*}

Figure\,\ref{fig:X-Rpol} presents the superposition of the X-ray contours on the polarized radio intensity map of the Southern lobe in \cen. As shown, some of the bright X-ray features classified before as `possibly related to the lobe' do indeed coincide with the polarized radio structures. The correlation is particularly evident in the case of the bright spots 7 and 8, located almost exactly at the position of the two ``holes'' within a single bright radio filament. This kind of morphology reminds us of the structures observed around the Galactic Center region, consisting of highly polarized radio filaments \citep{Lang99,LaRosa00,Nord04,Zadeh04} interacting with molecular clouds \citep{Staguhn98} and coinciding in many cases with compact X-ray features of unknown origin \citep[see also \S\,\ref{sec:discussion}]{Sakano03,Lu08,Johnson09}.

If the two ``on'' regions targeted with \suz\ are representative of
the rest of the halo, one may estimate the expected total
$0.5-10$\,keV flux and the total $0.5-10$\,keV luminosity of the
entire population of the X-ray spots `possibly related to the radio
structure' (with src\,3 excluded) to be $F_{\rm X,\,tot} \sim F_{\rm
  X} \times 13.2\,{\rm deg^2}/0.18\,{\rm deg^2} \sim 3 \times
10^{-11}$\,\Su, and $L_{\rm X,\,tot} \simeq 4 \pi D^2 \, F_{\rm
  X,\,tot} \sim 5 \times 10^{40}$\,\Lu, respectively, where
$13.2$\,deg$^2$ is the total angular extension of the \cen\ giant
lobes on the sky \citep[following][]{Hardcastle09}. These are
non-negligible values indeed, taking into account that the single
previously available SAS3 upper limit for the X-ray emission of the
halo is of the same order \citep{Marshall81}. Even more importantly,
the total expected non-thermal (inverse-Compton) flux of the giant
lobes within the $0.5-10$\,keV range, as derived from the broad-band
modeling presented in \citet{LAT10}, is $\simeq 1.2 \times
10^{-11}$\,\Su, which is formally below the expected summed
contribution from the spots. We also note that the total flux of the
\emph{diffuse} thermal emission detected in our \suz\ observations
within the \cen\ lobes is similarly of the same order of magnitude
(see \S\,\ref{sec:diffuse} below). It should be kept in mind, however,
that even if the discussed spots are indeed related to the lobes,
their distribution within the whole giant halo may not be uniform, as
the two analyzed \suz\ pointings --- which targeted small but
particularly radio-bright part of the Southern lobe --- may not be
representative of the rest of the extended structure.
Regardless, one may conclude that the total X-ray emission of giant
lobes in systems like \cen\ is a complex mixture of the diffuse
thermal and non-thermal components, \emph{plus} a prominent
contribution from compact X-ray features. Only with the broad energy
coverage and suitable combination of sensitivity and resolution of
\suz\ could these components be disentangled, as otherwise they would
form a single emission continuum.

Finally, let us comment on the three X-ray spots coinciding with
bright stars, and as such classified before as unrelated to the lobes.
Their XIS spectra, extracted in the same way as in the cases of src\,3
and 5--8 discussed above, did not give good fits with the \texttt{PL}
model (with $N_{\rm H}$ set free), or the absorbed \texttt{APEC} model
assuming an abundance of 0.3 solar or higher. Only with the abundance
in the \texttt{APEC} model set free and zero absorption could
reasonable fits be obtained, returning $Z < \, Z_{\odot}$ and $kT
\lesssim 1$\,keV. Low abundance in fits of this kind has been shown
often to be an artifact resulting from fitting to a mixed-temperature
plasma \citep[see the related discussion in][]{Kim12}. And, indeed, the
X-ray brightest spot coinciding with stars HD\,116099/TYC\,8248-981-1
could formally be fitted well with a two-component \texttt{APEC}
model. As noted above, this difference in the X-ray spectrum supports
our belief that these sources are unrelated to the lobes.

\subsection{Diffuse Emission Component}
\label{sec:diffuse}

In the previous sections we focused on compact X-ray features detected
in the \suz\ ``on'' pointings ``Lobe\,1'' and ``Lobe\,2'' centered on
the Southern lobe in the \cen\ radio galaxy. In this section we
present a detailed analysis of the diffuse emission component filling
the whole FOV of the XIS instrument, which remains \emph{after
  removing all the compact X-ray features} including src\,1--8, three
spots coinciding with bright optical stars, as well as spurious
sources due to instrumental artifacts. The diffuse counts extracted
from the entire field of view of both CCDs were analyzed using
standard methods, utilizing the energy range $0.6-7.0$\,keV for XIS0 and XIS3
and $0.5-6.0$\,keV for XIS1 (to avoid large errors in the
extrapolation of contamination files), and ignoring the range
$1.8-1.9$\,keV due to calibration uncertainty of Si K-edge. The ARF
files were produced assuming that the diffuse emission is distributed
uniformly within circular regions with 20\am\ radii (giving the ARF
area of 0.35\,deg$^2$) using \texttt{xissimarfgen} and new
contamination files\footnote{Contamination files
  \texttt{ae\_xi?\_contami\_edm\_20110927.fits} taken from
  \texttt{http://space.mit.edu/XIS/monitor/contam/}}. The background
component was estimated from the two shorter ``off'' \suz\ exposures
``Lobe\,3'' and ``Lobe\,4'' targeting the blank sky outside of (but
close to) the \cen\ Southern lobe (see the last two raws of
Table\,\ref{tab:obslog}), again after removing all the prominent
compact features.

\begin{figure}[!t]
\begin{center}
\includegraphics[width=\columnwidth]{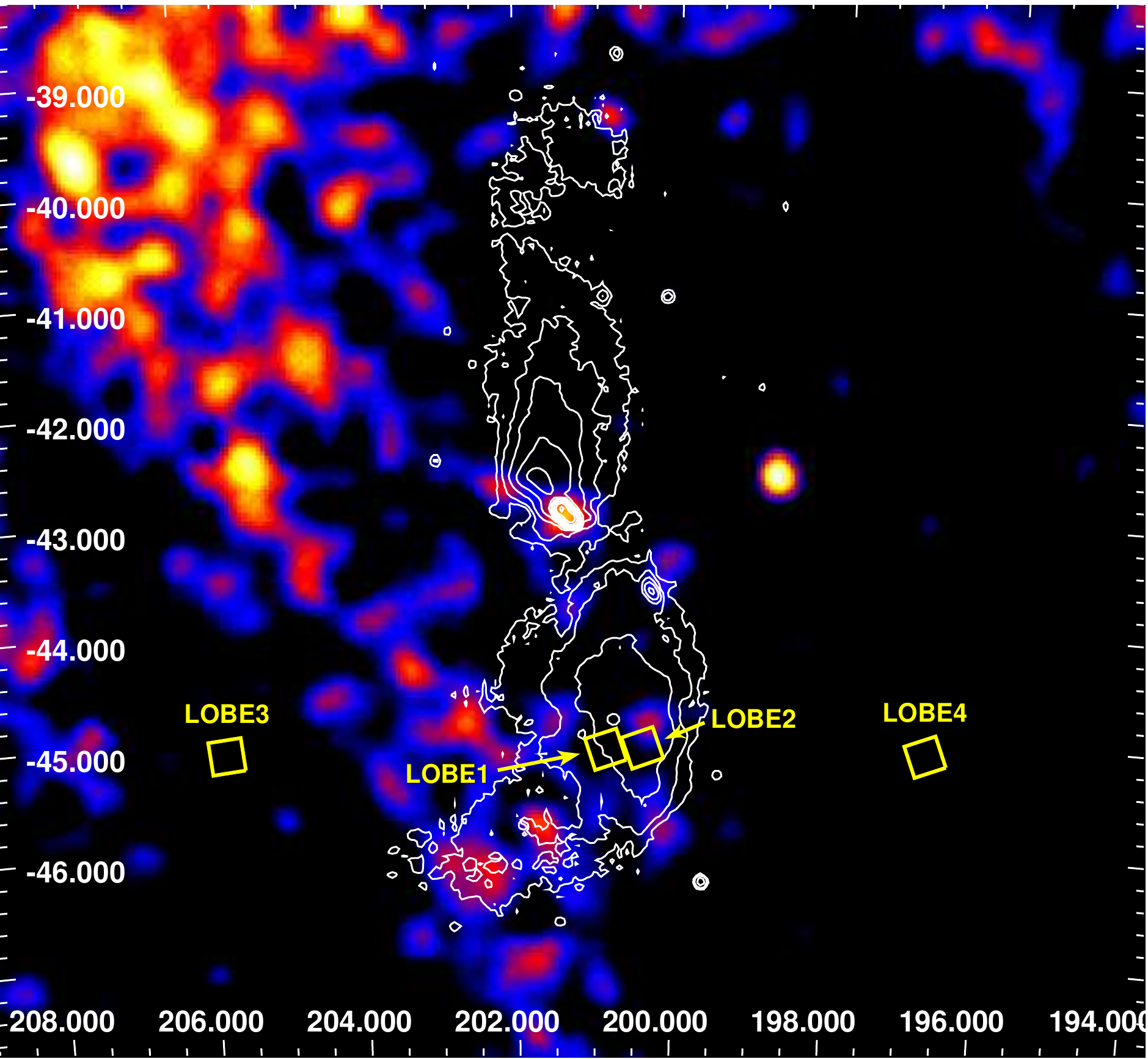}
\caption{\small 
Broad-band \textit{ROSAT} count map smoothed with a
Gaussian with $\sigma=7$ pixels and centered on the position of the
\cen\ radio galaxy. The white contours correspond to the 4.75\,GHz
Parkes image of the giant structure \citep{Junkes93}, begining at
0.1\,Jy/beam and increasing by a factor of two. The positions of all
the \suz\ pointings analyzed in this paper (see
Table\,\ref{tab:obslog}) are represented with yellow squares and
labelled with their observation number.
}{\label{fig:rosat}}
\end{center}
\end{figure}

\begin{figure}[!th]
\begin{center}
\includegraphics[width=\columnwidth]{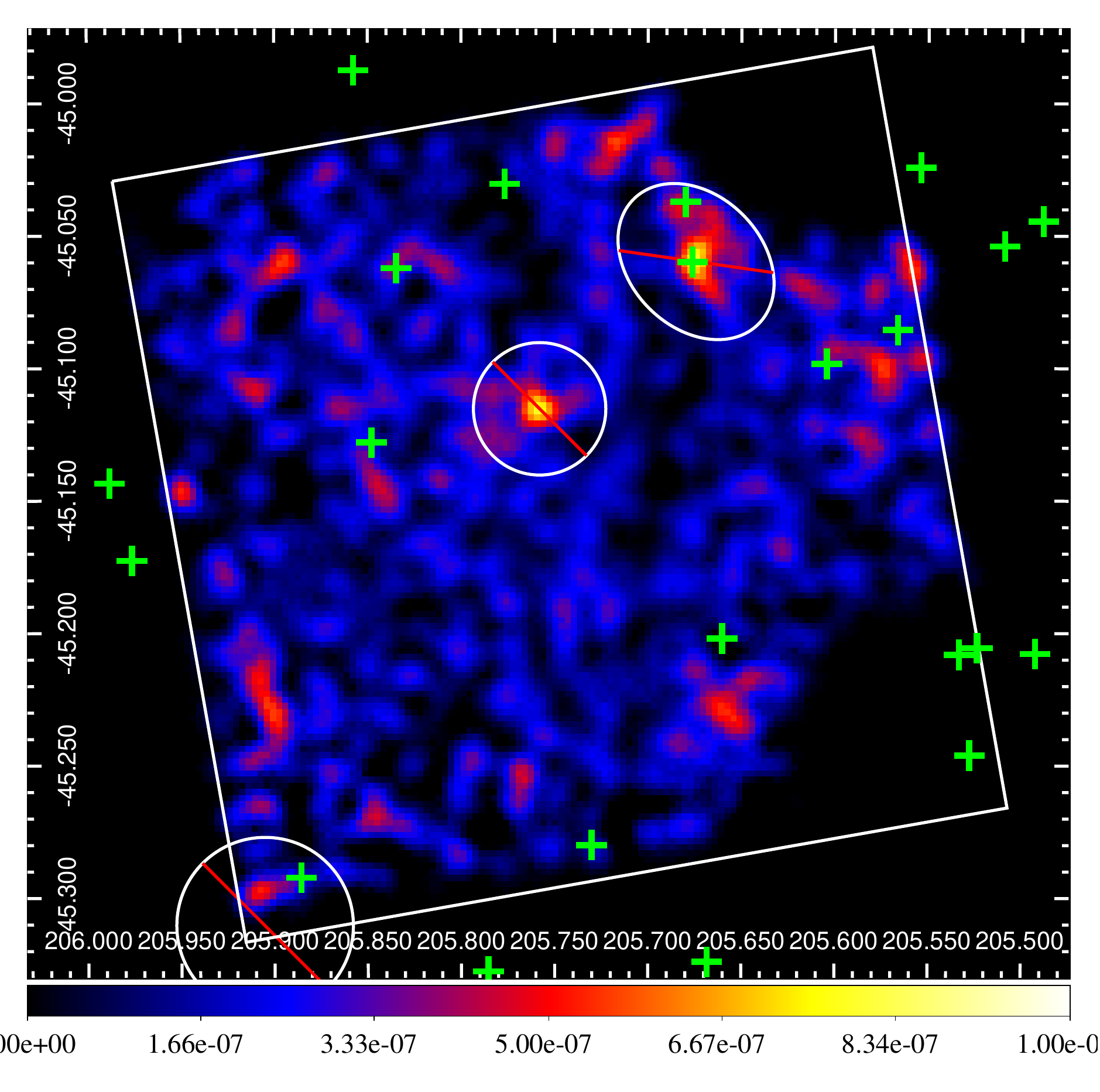}
\includegraphics[width=\columnwidth]{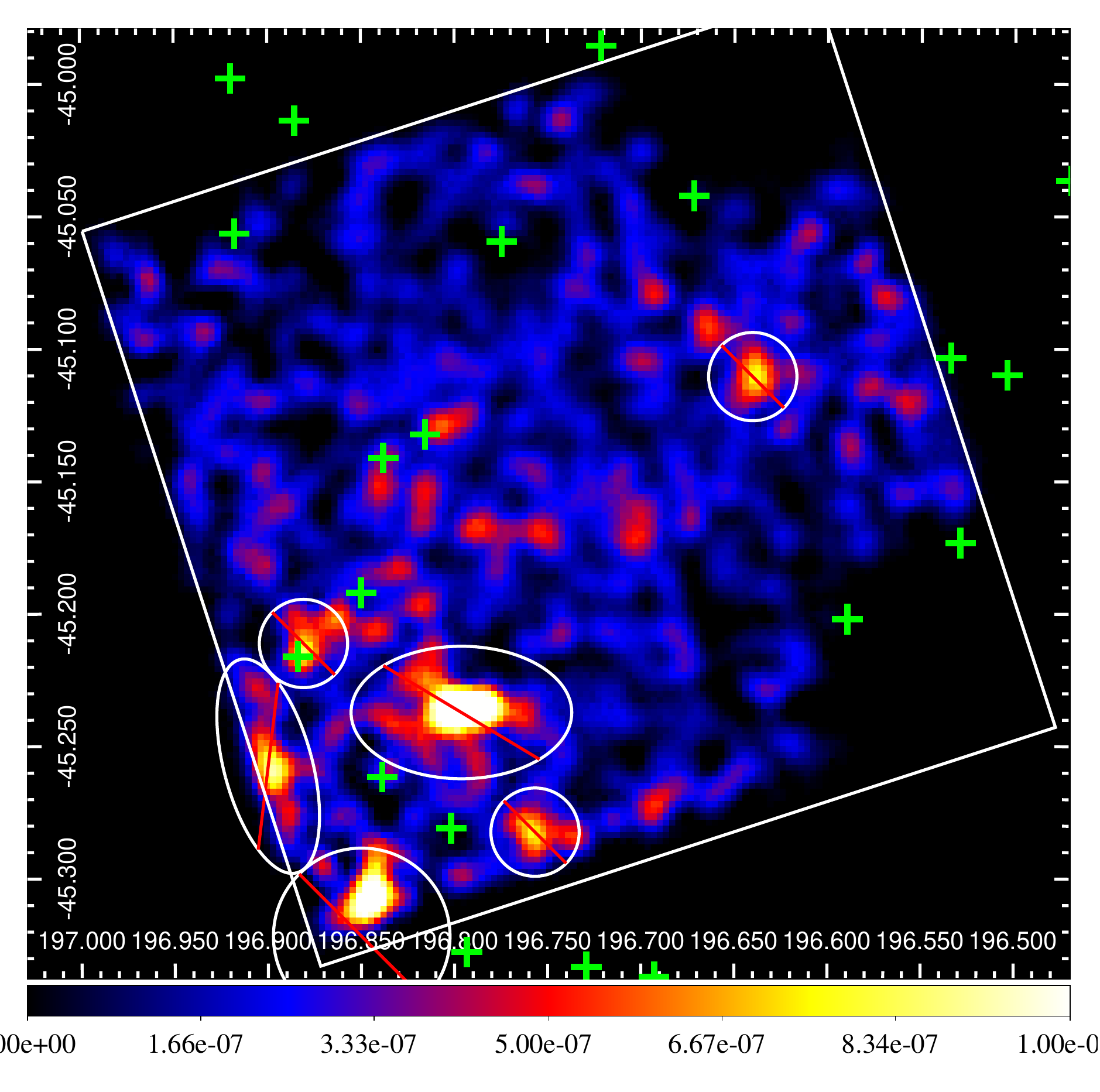}
\caption{\small 
Vignetting-corrected and NXB-subtracted XIS0+3 image of the ``off'' pointings (``Lobe 3'' and ``Lobe 4'' in the upper and lower panels, respectively; see Table\,\ref{tab:obslog}), with optical sources from the SIMBAD database denoted by green crosses. White circles and ellipses mark the positions of compact X-ray features which are removed from the extraction area used for the diffuse counts.
}{\label{fig:off}}
\end{center}
\end{figure}

Figure\,\ref{fig:rosat} presents the large-scale and broad-band
\textit{ROSAT} count map centered on the position of \cen, together
with the 4.75\,GHz contour map of the giant lobes \citep{Junkes93};
the positions of all the \suz\ pointings (see Table\,\ref{tab:obslog})
are marked with yellow squares. The XIS0+3 images of the ``off''
pointings, corresponding to the photon energy range $0.5-10$\,keV, are
shown in Figure\,\ref{fig:off}, with optical sources
from the SIMBAD database denoted by green crosses. In the figure,
white circles and ellipses mark the positions of compact X-ray
features (astrophysical sources or instrument artifacts in the
bottom-left corners of the two FOVs) which are removed from the
extraction areas used to measure spectra for the diffuse counts. We
note that the most prominent X-ray feature in the ``Lobe 4'' region
(lower panel in Figure\,\ref{fig:off}), with $0.5-10$\,keV flux
comparable to the those of src\,1--8 analyzed previously, coincides
with a strong ($\sim 10$\,mJy) radio source SUMSS\,J130713$-$451427.

First, we fit all four \suz\ pointings using \texttt{XSPEC} with the
same model \texttt{MEKAL+wabs*(MEKAL+PL)} including an unabsorbed
thermal component representing Local Bubble (LB) emission, an absorbed
thermal component representing Galactic Halo (GH) emission, and an
absorbed power-law component corresponding to the Cosmic X-ray
Background (CXB) radiation \citep[see, e.g.,][]{Yuasa09}. The neutral
hydrogen column density was fixed as before to the Galactic value in
the direction of the source, $N_{\rm H,\,Gal} = 0.7 \times
10^{21}$\,cm$^{-2}$, the photon index for the CXB component was fixed
at $\Gamma_{\rm CXB} = 1.41$, and the abundance in the \texttt{MEKAL}
models was fixed at the solar value. The results of the fitting are
summarized in Table\,\ref{tab:diff1}. As shown, while the LB and CXB
components are reproduced correctly in all four cases, an excess
emission in the ``on'' pointings seems to be present when compared
with \emph{both} ``off'' pointings, manifesting itself in the
increased (by a factor of two) normalization of the GH component in
the ``Lobe\,1'' and ``Lobe\,2'' regions. This excess may represent a
large-scale fluctuation in the GH emission, \emph{or} an additional
X-ray component originating in the \cen\ system. The same result is
obtained using the \texttt{APEC} plasma model instead of
\texttt{MEKAL}, along with an alternative parametrization of the
Galactic foreground and a complementary fitting procedure, as
summarized in the Appendix.

\begin{figure}[!th]
\begin{center}
\includegraphics[width=3.5in]{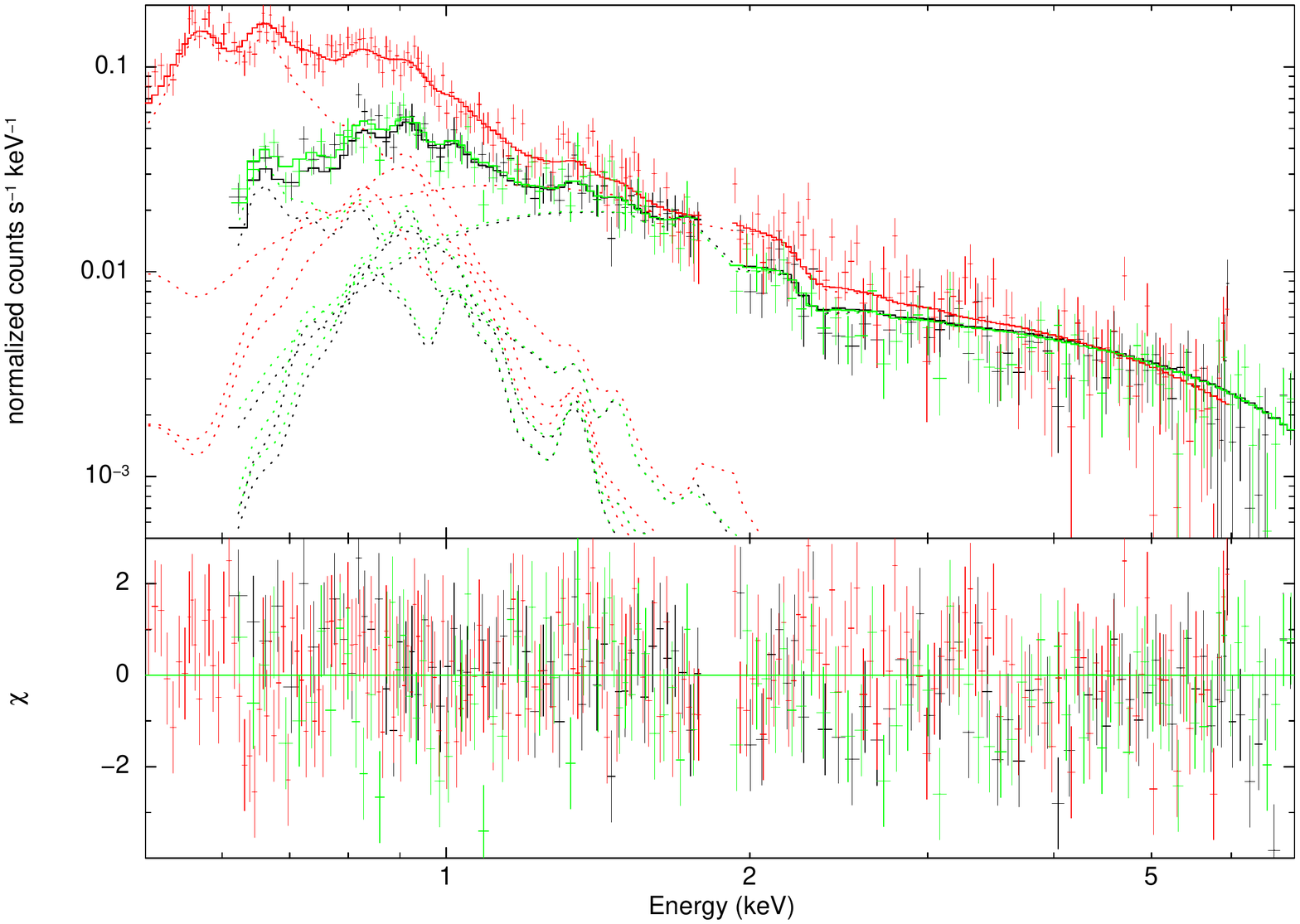}
\includegraphics[width=3.5in]{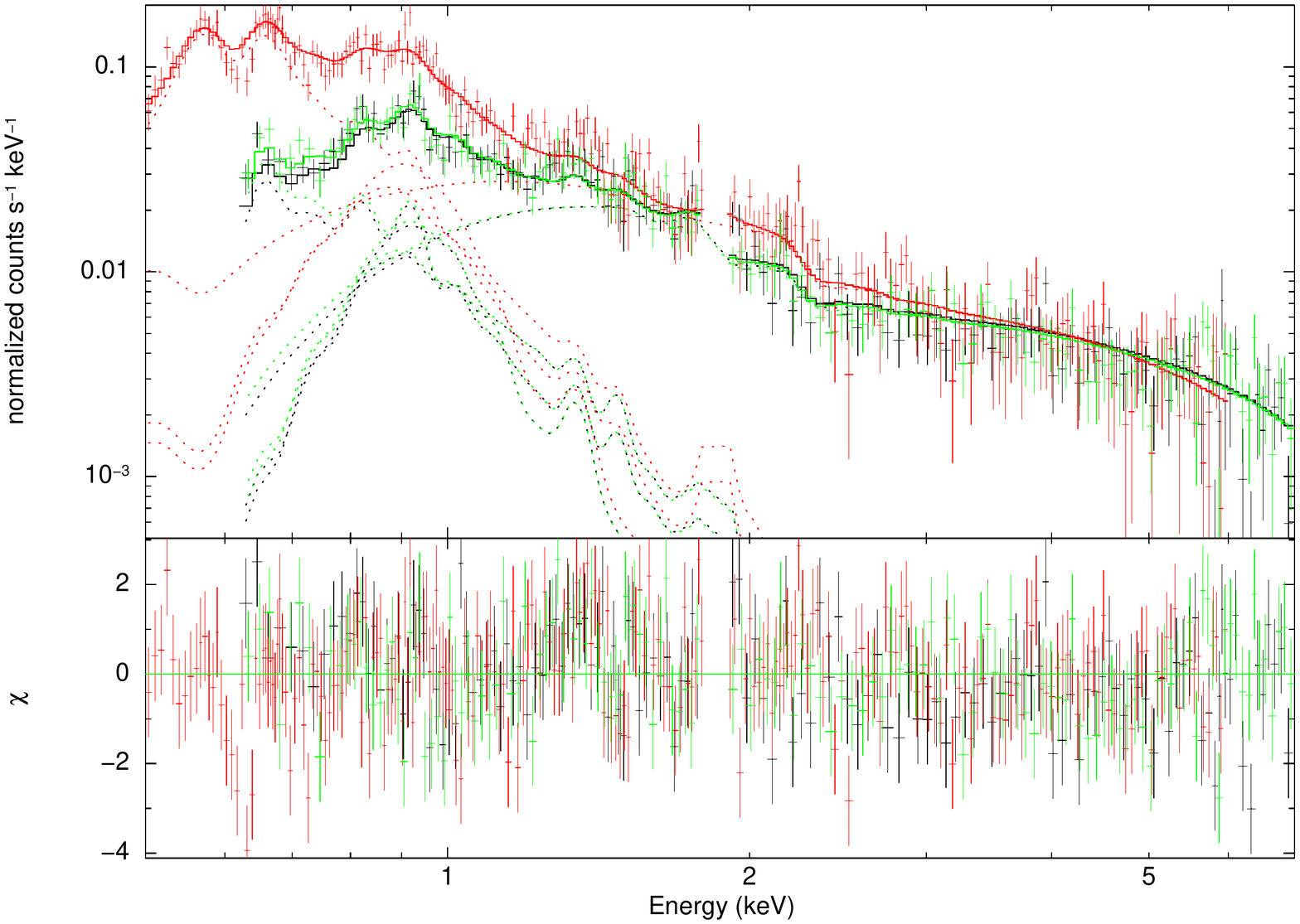}
\caption{\small 
The XIS spectra of the diffuse emission component detected in the ``Lobe\,1'' and ``Lobe\,2'' regions (upper and lower panels, respectively), together with the \texttt{MEKAL+wabs*(MEKAL+PL+MEKAL)} model curves corresponding to the $Z=0.3 \, Z_{\odot}$ fit described in \S\,\ref{sec:diffuse}. XIS0 data/fit are shown in black, XIS1 in red, and XIS3 in green.
}{\label{fig:diffuse}}
\end{center}
\end{figure}

Next, we fit the ``Lobe\,1'' and ``Lobe\,2'' regions assuming the
model \texttt{MEKAL+wabs*(MEKAL+PL+MEKAL)}, with the parameters
(temperatures and normalizations) of the LB+GH+CXB background emission
fixed as derived previously based on the analysis of the ``Lobe\,3''
region (see Table\,\ref{tab:diff1}), and an additional absorbed
thermal component representing the soft excess seen at the position of
the lobes. The results are summarized in Table\,\ref{tab:diff2}. The
same procedure using ``Lobe\,4'' instead of ``Lobe\,3'' background
parameters resulted in a rather poor description of the data (large
values of reduced $\chi^2$), because of a larger number of point
sources present in the ``Lobe\,4'' pointing which had to be removed
before the extraction of the diffuse counts, leading to a poorer
background determination. We repeated the fits assuming an abundance
for the additional thermal component of $Z/Z_{\odot}=0.1$, 0.3 and 1.
As expected, the derived model parameters are not sensitive to the
particular value of the metallicity assumed, except for normalization,
which roughly anti-correlates with $Z$. We note that in the case where
the soft excess is related to a large-scale fluctuation of the GH
emission, solar abundance should be expected. However, in the case
where the soft excess is related to the \cen\ structure, the abundance
might be expected either to be sub-solar, if the X-ray emitting gas is representative of the intergalactic medium interacting with the radio-emitting outflow \citep[see, e.g.,][]{Siemiginowska12}, or even super-solar, if it is related to the matter uplifted by the expanding radio bubble from the \cen\ host galaxy \citep[e.g.,][]{Simionescu08}. The corresponding XIS spectra of the diffuse emission detected in the ``Lobe\,1'' and ``Lobe\,2'' regions are presented in Figure\,\ref{fig:diffuse}, together with the model curves ($Z=0.3 \, Z_{\odot}$ fit). 

To sum up, we conclude that the model we have applied provides a
reasonable representation of the data, implying a relatively prominent
soft X-ray excess at the position of the \cen\ giant lobes probed with
\suz, which is best described as thermal emission from hot gas with a
temperature of $kT \sim 0.5$\,keV. As already noted above, it may
originate in the \cen\ system, or instead may be due to some
large-scale fluctuation in the foreground GH emission. The latter
option cannot be ruled out with the available data (cf. the \textit{ROSAT} 
map in Figure\,\ref{fig:rosat}).
However, in the discussion that follows (\S\,\ref{sec:diff-origin}) we
investigate the former possibility, arguing that the soft X-ray
emitting gas responsible for the observed excess is consistent with
diffuse matter filling the lobe and mixed with the non-thermal plasma,
or alternatively with a condensation of the intergalactic medium
around the edges of the expanding radio structure. The estimated
$0.5-2$\,keV absorbed flux of this gas is roughly $\simeq 6 \times
10^{-13}$\,\Su/0.35\,deg$^2$, meaning $\sim 2 \times
10^{-11}$\,\Su\ for the entire halo ($\simeq 13.2$\,deg$^2$), assuming
that the ``Lobe\,1'' and ``Lobe\,2'' regions are indeed representative
for the entire structure.

\begin{table*}[!th]
{\footnotesize
\noindent
{\caption[] {\label{tab:diff1} Modeling Results for the Diffuse Emission}}
\begin{center}
\begin{tabular}{cccccccccc}
\hline\hline
\\
region & $kT_{\rm LB}$ & Norm$_{\rm LB}$ & $kT_{\rm GH}$ & Norm$_{\rm GH}$ &  Norm$_{\rm CXB}$ & $F_{\rm 0.5-2\,keV}^{\rm abs}$ & $F_{\rm 2-10\,keV}^{\rm abs}$ & $Z/Z_{\odot}$ & red.$\chi^2$/dof\\
(1) & (2) & (3) & (4) & (5) & (6) & (7) & (8) & (9) & (10) \\
\\
\hline
\\
Lobe\,1 (``on'') & $0.19\pm 0.01$ & $3.53^{+0.20}_{-0.20}$ & $0.60^{+0.03}_{-0.03}$ & $7.42^{+0.79}_{-0.82}$  & $0.97^{+0.03}_{-0.02}$ & $7.88$ & $6.32$ & 1.0$^f$ & 1.16/441 \\
Lobe\,2 (``on'') & $0.20\pm 0.01$ & $3.22^{+0.18}_{-0.17}$ & $0.66^{+0.02}_{-0.03}$ & $7.38^{+0.68}_{-0.69}$  & $1.01^{+0.03}_{-0.02}$ & $7.67$ & $6.59$ & 1.0$^f$ & 1.13/512 \\
Lobe\,3 (``off'') & $0.20\pm 0.01$ & $3.30^{+0.33}_{-0.32}$ & $0.70^{+0.12}_{-0.09}$ & $3.79^{+1.28}_{-1.16}$  & $1.04^{+0.05}_{-0.05}$ & $7.19$ & $6.73$ & 1.0$^f$ & 1.30/131 \\
Lobe\,4 (``off'') & $0.19\pm 0.01$ & $2.34^{+0.33}_{-0.31}$ & $0.80^{+0.16}_{-0.13}$ & $3.45^{+0.88}_{-0.84}$  & $1.18^{+0.05}_{-0.06}$ & $5.87$ & $7.63$ & 1.0$^f$ & 0.87/123 \\
\\
\hline\hline
\end{tabular}
\end{center}
(1) Region fitted; (2) temperature of the LB component in the keV units; (3) \texttt{MEKAL} normalization of the LB component $\times 10^{-3}$; (4) temperature of the GH component in the keV units; (5) \texttt{MEKAL} normalization of the GH component $\times 10^{-4}$; (6) normalization of the CXB component $\times 10^{-3}$; (7) absorbed $0.5-2$\,keV flux in the units of $10^{-12}$\,\Su/0.35\,deg$^2$; (8) absorbed $2-10$\,keV flux in the units of $10^{-12}$\,\Su/0.35\,deg$^2$; (9) fixed (``$f$'') abundance; (10) reduced $\chi^2$ value/degree of freedom.
}
\end{table*}

\begin{table}[h]
{\footnotesize
\noindent
{\caption[] {\label{tab:diff2} Modeling Results for the Excess Diffuse Emission}}
\begin{center}
\begin{tabular}{ccccc}
\hline\hline
\\
$Z/Z_{\odot}$ & $kT$ & Norm & $F_{\rm 0.5-2\,keV}^{\rm abs}$ & red.$\chi^2$/dof\\
(1) & (2) & (3) & (4) & (5) \\
\\
\hline
{\bf Lobe\,1} & & & & \\
$1.0^f$ & $0.46^{+0.08}_{-0.11}$ & $3.11^{+1.11}_{-0.53}$ & 5.84  & 1.20/444 \\
$0.3^f$ & $0.45^{+0.08}_{-0.10}$ & $9.76^{+3.80}_{-1.88}$ & 6.12  & 1.20/444 \\
$0.1^f$ & $0.39^{+0.09}_{-0.07}$ & $27.3^{+10.4}_{-7.2}$ & 6.72  & 1.20/444 \\
\\
\hline
{\bf Lobe\,2} & & & & \\
$1.0^f$ & $0.64^{+0.05}_{-0.05}$ & $2.91^{+0.34}_{-0.35}$ & 5.83  & 1.13/515 \\
$0.3^f$ & $0.64^{+0.05}_{-0.05}$ & $8.59^{+1.03}_{-1.03}$ & 6.20  & 1.13/515 \\
$0.1^f$ & $0.62^{+0.06}_{-0.06}$ & $19.1^{+2.61}_{-2.49}$ & 6.72  & 1.13/515 \\
\\
\hline\hline
\end{tabular}
\end{center}
(1) The assumed abundance $Z$; (2) temperature
of the additional thermal component in the keV units; (3) the \texttt{MEKAL} normalization of the additional thermal component $\times 10^{-4}$; (4) absorbed $0.5-2$\,keV flux in the units of $10^{-13}$\,\Su/0.35\,deg$^2$; (5) reduced $\chi^2$ value/degree of freedom.
}
\end{table}

We finally note that the analysis of the data collected with the Hard
X-ray Detector (PIN instrument) on board \suz\ \citep{Takahashi07}
during the exposure that we carried out reveals no excess emission at
the position of the ``Lobe\,1'' and ``Lobe\,2'' over the background.
The implied $10-50$\,keV upper limit of $5 \times
10^{-12}$\,\Su/0.25\,deg$^2$, or $F_{\rm 10-50\,keV} < 3 \times
10^{-10}$\,\Su\ for the giant halo as a whole, is about one order of
magnitude above the non-thermal diffuse flux in the $10-50$\,keV range
expected from the broad-band modeling presented in \citet{LAT10},
namely $\sim 2 \times 10^{-11}$\,\Su. At lower photon energies, it is
the level of the CXB component of the spectrum that provides the
relevant upper limit for the non-thermal diffuse emission of the
lobes. In detail, the total absorbed $2-10$\,keV energy flux
detected in the ``on'' pointings is $\simeq 6.5 \times
10^{-12}$\,\Su/0.35\,deg$^2$ (see Table\,\ref{tab:diff1}), basically
as expected for the CXB surface brightness. This implies that any
diffuse emission filling the halo should not exceed $\sim 10\%$ of
this value. Meanwhile, the $2-10$\,keV energy flux expected for the
Southern lobe of \cen\ based on the \citet{LAT10} analysis is on
average $\simeq 1.5 \times 10^{-13}$\,\Su/0.35\,deg$^2$, i.e. about a
factor of four below the formal upper limit of $10\%$ of the CXB
level.

\section{Discussion}
\label{sec:discussion}

\subsection{Origin of the Diffuse Emission}
\label{sec:diff-origin}

The results of the analysis of the \suz\ data regarding the diffuse
emission positioned at the part of the Southern giant lobe
in the \cen\ radio galaxy, as summarized above, suggest a presence of
an excess component best modeled as thermal emission from a hot gas
with temperature $kT \simeq 0.5$\,keV. Assuming this excess originates
within the \cen\ halo rather than in the Galactic foreground, the
returned \texttt{MEKAL} normalization (see Table\,\ref{tab:diff2}),
\begin{equation}
{\rm Norm} = \frac{10^{-14}}{4 \pi D^2} \times \left(\frac{n_g}{{\rm cm^{-3}}}\right)^2 \times \left(\frac{\rm ARF}{{\rm cm^{2}}}\right) \times  \left(\frac{\ell}{{\rm cm}}\right) \, ,
\end{equation}
\noindent
where ${\rm ARF} = 0.35$\,deg$^2$ with the conversion scale
$1.08$\,kpc/arcmin and $\ell$ is the third dimension of the emitting
region, implies a number density for the X-ray emitting gas of $n_g
\simeq (0.9-2.5) \times 10^{-4}$\,cm$^{-3}$, where we assume
abundances within the range $10\%-100\%$ solar, and $\ell = 150$\,kpc
as appropriate for gas that uniformly fills the entire Southern lobe
\citep[see][]{Hardcastle09}. Such a value for the density may be
considered as relatively high, implying that the total mass of the
thermal gas contained within the entire halo is of the order of
$\bar{n}_g m_p V_{\ell} \sim 10^{10}\, M_{\odot}$ for the mean number
density $\bar{n}_g \simeq n_g \sim 10^{-4}$\,cm$^{-3}$ within the
lobes' volume $V_{\ell} \simeq 2 \times 10^{71}$\,cm$^{-3}$. Yet if
such a large amount of hot gas is indeed mixed with the diffuse
non-thermal plasma, composed of a magnetic field with the
volume-averaged field strength $\bar{B} \simeq 1$\,$\mu$G roughly in
pressure equipartition with the radio-emitting ultrarelativistic
electrons --- namely, $\langle U_B \rangle \equiv {\bar{B}}^2 / 8 \pi
\simeq \langle p_{e\pm} \rangle \equiv \langle U_{e\pm} \rangle/3 $
\citep{LAT10} --- then the lobes' thermal pressure $\langle p_g
\rangle = \bar{n}_g \, k\bar{T} \simeq 8 \times 10^{-14}$\,\Uu\ is
almost exactly equal to the non-thermal one $\langle p_{e\pm} \rangle+
\langle U_B \rangle \simeq 8 \times 10^{-14}$\,\Uu. In other words,
the sound velocity in the system $\bar{c}_s = (5 \, k\bar{T} / 3 \,
m_p)^{1/2} \simeq 3 \times 10^7$\,cm\,s$^{-1}$ is then approximately
equal to the Alfven velocity $\bar{v}_A = \bar{B} / (4 \pi \, m_p
\bar{n}_g)^{1/2} \simeq 2 \times 10^7$\,cm\,s$^{-1}$, or equivalently
the plasma parameter $\beta \equiv \langle p_g \rangle / \langle U_B
\rangle \simeq (\bar{c}_s / \bar{v}_A)^2 \sim 1$. Pressure
equilibration between different plasma species in the giant lobes of
\cen\ is a meaningful characteristic, reminding us again of the
ISM within the Galactic disk.

The numbers derived above challenge at the same time, at least to some
extent, the alternative scenario for the soft X-ray excess related to
the \cen\ system, and involving a condensation of the hot
intergalactic medium around the edges of the expanding radio
structure. That is because in such a case the third dimension of the
emitting volume is likely to be smaller than the scale of the
lobes\footnote{We note, however, that in the case of very light
  relativistic jets evolving in a stratified hot thermal atmosphere,
  it is possible to find a set of the model parameters for which the
  spatial scale of the gas compressed and heated by the bow shock
  driven by the expanding cocoon is comparable to the scale of the
  radio lobes (Hardcastle \& Krause, in prep.).}, $\ell \ll 150$\,kpc,
resulting in an increased gas number density $n_g \propto \ell^{-1/2}$
and therefore in an increased gaseous pressure $p_g \propto n_g$
largely exceeding $\langle p_{e\pm} \rangle + \langle U_B \rangle$.
That is to say, this alternative scenario does not give
self-consistent values for the model parameters, \emph{unless} an
additional pressure support is assumed for the lobes, for example due
to ultrarelativistic protons injected by the jets into the expanding
cocoon \citep[see, e.g.,][]{Ostrowski00}. We also note in this context
that the gas density in a typical poor group of galaxies at the
distance $r$ from the group center, approximated by the standard
profile
\begin{equation}
n_{\rm igm}\!(r) \simeq n_{0} \times \left( {r \over a_{0}} \right)^{- b}
\end{equation}
\noindent
with $r > a_{0} \simeq 10$\,kpc, $b \simeq 1.5$, and $n_{0} \lesssim
10^{-2}$\,cm$^{-3}$, yields the expected $n_{\rm igm}\!(150\,{\rm
  kpc}) \lesssim 10^{-4}$\,cm$^{-3}$; the relatively large scatter in
the values of the parameters $a_0$, $b$ and $n_0$ found for various systems should be kept in mind \citep[e.g.;][]{Mulchaey98,Sun12}

In general, the extended lobes in radio galaxies and quasars were
previously believed to be largely devoid of thermal matter, and to
consist solely of relativistic particles and magnetic field deposited
by a pair of jets over the entire lifetime of a source
\citep[e.g.,][]{Scheuer74,Begelman89,Kino12}. More recently, this
somewhat simplistic view has changed, as several observational
findings triggered the discussion on possibly significant thermal
content of the lobes in different systems. For example, observations
of so-called `double-double radio galaxies', i.e. the systems with
inner/younger radio lobes evolving within and aligned with the
outer/older lobes \citep{Schoenmakers00,Saikia09}, suggested a mass
density within the outer lobes in excess of that expected from a
pure jet deposition. \citet{Kaiser00} discussed in this context
various mechanisms enabling the extended radio lobes to be enriched by
the additional thermal matter, e.g., via uplifting of some part of the
interstellar medium from the galaxy host by the expanding outflow, as
in fact observed recently in a few cases \citep{Simionescu08}, or via
the entrainment of warm and dense gaseous clouds present in the
intergalactic medium by the lobes (followed by the clouds' destruction
and dispersion over the entire lobe volume on the time scales of about
$\sim 10$\,Myr). In addition, analysis of the pressure balance
between the radio lobes and the surrounding intergalactic or
intracluster medium for larger samples of objects led to the
conclusion that the dominant pressure support in the lobes, in the
case of the oldest and most evolved systems, cannot be entirely
explained by the radio-emitting electrons and magnetic field, but in
addition requires either thermal matter or a relativistic proton
population \citep{Hardcastle00,Dunn05,Croston08,Birzan08}. These are
all strong yet indirect indications for the presence of thermal matter
within the extended lobes of radio-loud AGN. Until now, no direct
evidence for such matter has been found in either X-ray or radio
studies \citep{Sanders07}.

Upper limits for the thermal gas density within the lobes of luminous
radio galaxies were previously derived from the apparent lack of any
signatures of internal depolarization of the radio emission, with
often claimed volume- and sample-averaged product $\langle n_g \times
B \rangle < 5 \times 10^{-3}$\,$\mu$G\,cm$^{-3}$ \citep{Garrington91},
meaning $\bar{n}_g < (1-10) \times 10^{-4}$\,cm$^{-3}$ for the
typically derived field strength range $3$\,$\mu$G\,$<B<30$\,$\mu$G
\citep[and references therein]{Kataoka05,Croston05,Isobe11}. In the
particular case of \cen, the analysis of the improved RM data
presented in the forthcoming paper \citet{Shane12} reveals, for the
very first time, strong indication for a positive detection of the
internal depolarization signal, with a corresponding gas density
$\bar{n}_g \sim 10^{-4}$\,cm$^{-3}$. This is in an excellent agreement
with the results presented here based on a completely different
dataset. Such an agreement is rather encouraging, keeping in mind that
the RM analysis by \citet{Shane12} should be considered to be more
robust method for the detection of the thermal content of the
\cen\ giant lobes, as it probes the entire volume of the halo and not
only a small part of them as in the case of the \suz\ data discussed here.

\subsection{Origin of Compact X-ray Features}
\label{sec:spot-origin}

In this section we return to the compact X-ray `spots' detected in our
\suz\ observations and discuss their origin in more detail. In the
estimates given below we evaluate plasma parameters corresponding to
different emission models for a single spot with radius $R \leq
1.5$\,\am\,$\simeq 1.6$\,kpc producing an $0.5-10$\,keV flux at the
level of $F_{\rm x} \simeq 10^{-13}$\,\Su, i.e. characterized by an
X-ray luminosity of, roughly, $L_{\rm x} \simeq 10^{38}$\,\Lu\ (these
parameters are roughly characteristic of the spots we observe; see
\S\,\ref{sec:spotmodel}).

\subsubsection{Bremsstrahlung Emission}

A thermal interpretation of the detected X-ray spots would imply the
presence of compact over-densities in the gas distribution within the
giant lobes of \cen, with the temperatures $kT \simeq 5$\,keV elevated
by an order of magnitude with respect to the thermal diffuse pool
characterized by the average $k\bar{T} \simeq 0.5$\,keV. Such
over-densities could, in principle, constitute manifestations of the
interactions of the magnetized radio-emitting outflow with
pre-existing inhomogeneities in the ambient medium, as in fact
observed in \cen\ on somewhat smaller scales, around the middle and
inner Northern lobes, much closer to the active nucleus and the host
galaxy \citep[see][]{Morganti10}. There, the performed infrared,
optical and X-ray studies indicate the presence of dust clouds,
filaments of young stars, or condensations of hot gas related to the
radio structure, and best explained in terms of interactions of the
expanding jets/lobes in the system with the ambient matter, resulting
in substantial heating of the plasma and the triggering of star formation
\citep[e.g.,][]{Kraft09,Auld12,Crockett12}. Similar phenomena are also
observed in other analogous radio galaxies, and hence seem generic for
nearby, low-power radio-loud AGN \citep[e.g.,][and references
  therein]{Siemiginowska12}.

In this context, radio lobes may be analogous to the Galactic Center
region, where the disturbed endpoints of prominent non-thermal radio
filaments are observed to coincide with molecular clumps, consistent
with the idea that magnetic energy is liberated as the large-scale
magnetic field, tangled by the motion of the clumps, reconnects in the
externally ionized surface layer of the advancing clumps
\citep[e.g.,][]{Staguhn98,Sofue05}. This type of morphology reminds us
to some extent of the complex structure in the Southern giant lobe of
\cen\ revealed by our new X-ray and radio maps, suggesting a scenario
in which the energy of the reconnecting magnetic field, released
primarily in the form of short-scale magnetic turbulence, is used to
heat the plasma of the interacting gaseous condensation, and to form
in this way the observed compact X-ray--emitting features, while also
depolarizing the synchrotron radio emission of the filaments (cf.
src\,7 and 8 in Figure\,\ref{fig:X-Rpol}).

In the framework of this model, the plasma energization timescale
involved has to be shorter than the lifetime of the system, $\tau_{\rm
  life} \sim 30$\,Myr \citep{Hardcastle09}, but not much shorter than
the Coulomb collision timescale of the heated electron population in
order to avoid any efficient acceleration of particles to higher
(ultrarelativistic) energies and hence the formation of an
energetically relevant non-thermal tail in the electron energy
distribution \citep[see the discussion in][]{Petrosian08}. Assuming
the `cold ambient plasma' regime for the Coulomb interactions, the
appropriate collision timescale can be evaluated to be
\begin{equation}
\tau_{\rm Coul} \simeq \frac{2 \, E_e \, v_e}{3 \sigma_T \, m_e c^4 \, n_g \, \ln\!\Lambda} \sim  0.5 \, n_{-4}^{-1} \,{\rm Myr}
\end{equation}
for the Coulomb logarithm $\ln\!\Lambda \simeq 30$, electron kinetic energy $E_e = 0.01 \, m_e c^2$ with the corresponding electron velocity $v_e$, and the gas density $n_g \equiv n_{-4} \times 10^{-4}$\,cm$^{-3}$. It should be kept in mind, at the same time, that the `cold plasma regime' is not strictly valid for the electrons with $v_e$ close to the thermal velocities of the background particles, in which case the Coulomb collision timescale may be much longer than the one evaluated above \citep[see][]{Petrosian08}.

The problem with the above interpretation is however the energetics of the spots. The \texttt{APEC} normalization $\simeq (0.5-1) \times 10^{-4}$ derived for the analyzed src\,5--8 (see Table\,\ref{tab:fits}) implies gas number densities $n_g \sim (3-5) \times 10^{-3}$\,cm$^{-3}$ for the assumed abundance $30\%$ solar, the source extraction region $\pi \, R^2$ and the third dimension of the emitting volume $2 \times R$. In the limiting case of negligible metallicity, the number density of the gas with a temperature $kT \simeq 5$\,keV needed to produce $10^{38}$\,\Lu\ of thermal X-rays within a spot volume $V = (4/3) \pi R^3$ reads as
\begin{equation}
n_g \simeq \sqrt{\frac{3 m_e^{3/2} c^3 h \, (kT)^{1/2} \, L_{\rm x}}{32 \, (2 \pi)^{1/2} e^6 \, V \, f_{\rm TB}\!(T)}} \sim 6 \times 10^{-3} \, R_{1.5}^{-3/2} \, {\rm cm^{-3}} \, ,
\end{equation}
where $R_{1.5} \equiv R/1.5$\am, and
\begin{equation}
f_{\rm TB}\!(T) = \int_{\varepsilon_{\rm min}}^{\varepsilon_{max}}\!\! d\varepsilon \,\,\, \ln\!\left(\frac{4 \, kT}{1.78 \, \varepsilon}\right) \times \exp\!\left(-\frac{\varepsilon}{kT}\right)
\end{equation}
with $\varepsilon_{\rm min} = 0.5$\,keV and $\varepsilon_{\rm max} =
10$\,keV. This is about two orders of magnitude above the number
density of the diffuse thermal gas within the \cen\ giant lobes,
leading to a situation where the X-ray emitting clumps are
\emph{highly} over-pressured with respect to their surroundings, $n_g
kT \sim 5 \times 10^{-11} \, R_{1.5}^{-3/2}$\,\Uu\,$\gg \langle p_g
\rangle$. One could speculate, on the other hand, that the compact
features we observe are relatively short-lived, and that soon after
the onset of a violent reconnection event in an isolated region within
the lobe, followed by a rapid and efficient turbulent heating of the
plasma in a confined volume, a newly-formed over-pressured spot
disappears quickly on a sound-crossing timescale of, roughly, $\tau_s
\simeq R / c_s \leq 2 \, R_{1.5}$\,Myr. This, possibly, would
correspond to the situation of a rather efficient transfer of the
magnetic energy to the internal energy of the lobes' thermal plasma
during the lifetime of the source.

\subsubsection{Inverse-Compton Emission}

One possibility for production of non-thermal X-ray emission from the
spots is inverse-Compton (IC) up-scattering of Cosmic Microwave
Background (CMB) photons (the dominant radiation field at the
considered distance from the galactic host) by ultrarelativistic
electrons with Lorentz factors $\gamma \simeq \sqrt{\varepsilon_{\rm
    x} / \varepsilon_{\rm cmb}} \sim 10^3$, where $\varepsilon_{\rm x}
\simeq 1$\,keV and the characteristic energy of the CMB radiation is
$\varepsilon_{\rm cmb} \simeq 1$\,meV. The involved electron
population could originate, for example, from a
compression/re-acceleration of cosmic rays filling the giant lobes
\citep[and producing diffuse non-thermal emission at radio and X-ray
  frequencies as modeled in][]{Hardcastle09,LAT10} by kpc-scale shocks
formed around the reconnecting radio filaments.

Let us approximate the energy distribution of such a hypothetical additional electron population by an arbitrary form $dN_e(\gamma)/d\gamma \propto \gamma^{-3}$ between, say, $\gamma_{\rm min} = 10$ and $\gamma_{\rm max} = 10^3$, as very roughly justified by the X-ray photon indices emerging from the \texttt{PL} fits to the spectra of the spots ($\Gamma \sim 2$ on average, albeit with a large spread). With such an assumption, the pressure of the electron population involved in producing the observed X-ray emission is
\begin{equation}
p_{e\pm} \simeq \frac{m_e c \, \gamma_{\rm min}^{-1} \, L_{\rm x}}{4 \sigma_T  \, \ln\!\left(\frac{\gamma_{\rm max}}{\gamma_{\rm min}}\right) \, U_{\rm cmb} \, V} \sim 10^{-10} \, R_{1.5}^{-3} \, {\rm erg\,cm^{-3}} \, ,
\end{equation}
where $U_{\rm cmb} \simeq 4 \times 10^{-13}$\,\Uu\ is the energy
density of the CMB photon field, and we put $\int \gamma^2
dN_e\!(\gamma) / \int \gamma \, dN_e\!(\gamma) \simeq
\ln\!(\gamma_{\rm max}/\gamma_{\rm min})/ \gamma_{\rm min}^{-1}$.
Similarly to what we found in the case of a thermal model, this
pressure exceeds by orders of magnitude the total pressure of the
ambient plasma (especially for $R_{1.5} < 1$), which invalidates the
scenario drafted above, and hence rules out an IC/CMB origin for the
emission of the spots detected with \suz. The particular assumption
regarding the exact form of the electron energy distribution used in
the estimate above is not crucial in this respect, since even in the
extreme case of a mono-energetic population of radiating electrons
approximated by the delta function, $dN_e(\gamma)/d\gamma \propto
\delta(\gamma - \gamma_{\rm max})$, the corresponding electron
pressure would be reduced only by a factor of $\ln\!(\gamma_{\rm
  max}/\gamma_{\rm min})/ (\gamma_{\rm max}/\gamma_{\rm min}) \simeq
0.04$, i.e. would still be very high, $\gg \langle p_g \rangle$.

\subsubsection{Synchrotron Emission}

The other non-thermal scenario to consider for the discussed X-ray
features is the synchrotron radiation of very high energy
ultrarelativistic electrons. We note that synchrotron X-ray emission
is routinely detected in blazars of the `high frequency peaked BL Lac'
type \citep[e.g.,][]{Takahashi01}, young supernova remnants
\citep{Vink12}, kpc-scale jets in low-power radio galaxies
\citep{Harris06}, or pulsar wind nebulae \citep{Gaensler06}. In all
these system an efficient \emph{in situ} acceleration of the
X-ray--emitting electrons up to $\geq$\,TeV energies is required,
although the exact energy dissipation processes involved, which in
general are ill-understood, may be substantially different in
different classes of objects. The particular multiwavelength
morphology revealed by our maps, with some of the compact X-ray spots
coinciding with extended and highly polarized radio filaments, prompts
us to consider in this context the scenario where the magnetic energy
released at the locations of reconnecting magnetic tubes/radio
filaments is responsible for the confinement and acceleration of
relativistic particles via a Fermi process within distinct emission sites
\citep{Drake06,Hoshino12}.

Let us therefore consider this possibility more quantitatively, in the 
analogy to the turbulent acceleration
scenario developed in the context of solar flares \citep[see][and 
references therein]{Petrosian12}. For this we simply assume that within the
volume, which may be approximated as spherical with radius $R$,
small-scale magnetic turbulence is efficiently injected around the
reconnecting large-scale magnetic field $B$, so that the total
turbulent energy is roughly comparable to the magnetic energy,
$(\delta B / B)^2 \sim 1$. With such strong turbulence conditions, the
mean free path of \emph{ultrarelativistic} electron for particle-wave
interactions is expected to be of the order of the electron
gyroradius, $\lambda_{e\pm} \sim r_{e\pm} = \gamma \, m_e c^2 / e B
\ll R$, and the spatial diffusion coefficient is $\kappa_{e\pm} \sim
\lambda_{e\pm} \, c /3$. Hence the timescale for the particle escape
from the acceleration site $\tau_{\rm esc}\!(\gamma) \sim R^2 /
\kappa_{e\pm}$. The spatial diffusion of the electrons is obviously
accompanied by their diffusion in momentum space, leading to a net
acceleration \emph{at the maximum allowed rate} with the
characteristic timescale $\tau_{\rm acc}\!(\gamma) \sim \lambda_{e\pm}
c / v_{\rm sc}^2$, where $v_{\rm sc}$ is the velocity of the
interacting waves (Alfven or fast magnetosonic waves in the case of
ultrarelativistic electrons). The evolving electrons also undergo
radiative energy losses, dominated in our case by the synchrotron and
IC/CMB processes; the characteristic timescale for these can be
approximated as $\tau_{\rm rad}\!(\gamma) \sim 3 m_e c/4 \sigma_T
\gamma U_0$, where $U_0\!(\gamma) = U_B + U_{\rm cmb} \, f_{\rm
  KN}\!(\gamma)$, $U_B \equiv B^2/8 \pi$, and $f_{\rm KN}\!(\gamma) =
(1 + 4 \gamma \varepsilon_{\rm cmb} / m_e c^2)^{-3/2}$ stands for the
Klein-Nishina correction to the IC scattering cross section. 
The synchrotron scenario can therefore be considered as a viable option if
the maximum available (`equilibrium') electron energy $\gamma_{\rm
  eq}$ given by the condition $\tau_{\rm acc}\!(\gamma_{\rm eq}) \sim
\tau_{\rm rad}\!(\gamma_{\rm eq}) \ll \tau_{\rm esc}\!(\gamma_{\rm
  eq})$ is large enough to produce synchrotron photons with
$\varepsilon_{\rm x} = 3 h e B \gamma^2/ 4 \pi m_e c \sim 1$\,keV
energies. This is indeed possible to achieve for a magnetic field only
slightly larger than the volume-averaged value, $B \gtrsim
1$\,$\mu$G\,$\simeq \bar{B}$, as in fact would be expected for highly
polarized filaments \citep{Feain11}.

\begin{figure}[!t]
\begin{center}
\includegraphics[width=\columnwidth]{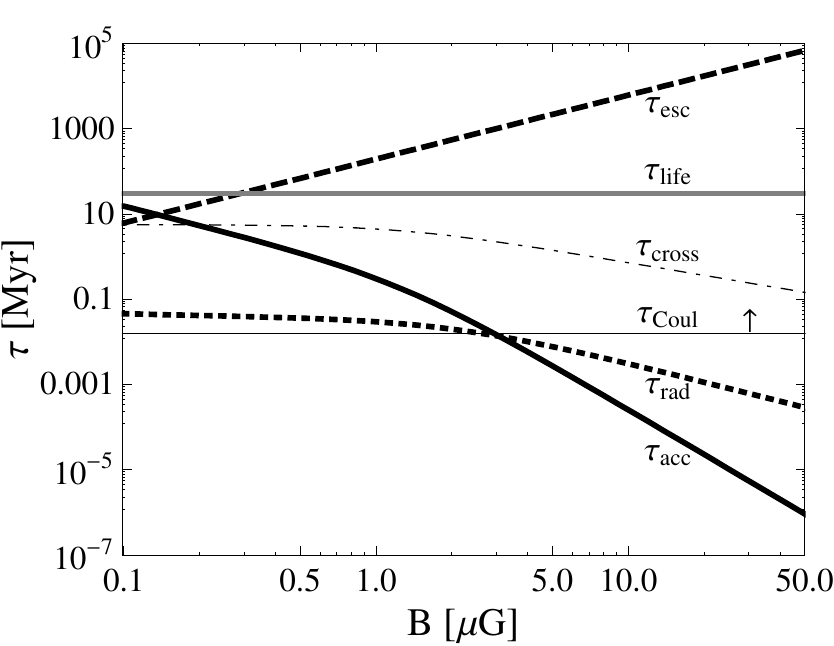}
\caption{\small 
Acceleration, cooling and escape timescales for ultrarelativistic
electrons emitting synchrotron X-rays, as functions of the magnetic
field $B$ within the considered emission site. The thick dotted curve
denotes the radiative loss timescale, thick solid curve corresponds to
the acceleration timescale, and the thick dashed curve illustrates the escape timescale for an emission region size $R=1.6$\,kpc. We also plot the age of the \cen\ giant lobes (horizontal gray line), the magnetosonic crossing timescale (thin dot-dashed curve), and the lower limit for the Coulomb collision timescale for the background thermal plasma (thin horizontal line).
}{\label{fig:timescales}}
\end{center}
\end{figure}

In Figure\,\ref{fig:timescales} we plot all the analyzed timescales
corresponding to $\gamma_{\rm eq}$ electrons emitting synchrotron
$\varepsilon_{\rm x} = 1$\,keV photons as functions of the magnetic
field $B$ within the considered region of radius $R$. The thick dotted
curve denotes the radiative loss timescale. As shown, synchrotron
losses dominate over the IC/CMB process for $B \gtrsim$ a few $\mu$G.
The thick solid curve in the figure corresponds to the acceleration
timescale with the velocities of the turbulent modes set to be $v_{\rm
  sc} = \sqrt{v_A^2 + c_s^2}$, where the Alfven velocity $v_A$ is
evaluated for a background gas density $\bar{n}_g =
10^{-4}$\,cm$^{-3}$, while the sound velocity $c_s$ assumes a
background gas temperature $k\bar{T} = 0.5$\,keV. It follows that the
Alfvenic acceleration dominates for $B >1$\,$\mu$G. The thick dashed
curve shows the escape timescale of the X-ray--emitting electrons,
$\tau_{\rm esc} > 100$\,Myr for $R =1.6$\,kpc and $B > 1$\,$\mu$G. For
comparison, we also plot the estimated lifetime\footnote{The lifetime
  estimates of \cite{Hardcastle09} are derived on the assumption of no
  efficient in situ particle acceleration in the lobe, and so the use
  of this lifetime estimate in this context is not strictly
  self-consistent; we retain it for simplicity, noting that
  significant particle acceleration would tend to artificially reduce
  lifetimes estimated with spectral ageing methods, and also that 
  in the framework of the discussed model very efficient electron 
  acceleration is restricted only to compact turbulent regions of the 
  lobes, which we identify as sites of violent reconnection of large-scale 
  magnetic tubes/radio filaments, and is not distributed within the whole 
  volume of the giant halo.} of the giant halo $\tau_{\rm
  life} \simeq 30$\,Myr (thick horizontal gray line), the magnetosonic
crossing timescale $\tau_{\rm cross} \simeq R/v_{\rm sc}$ with $R
=1.6$\,kpc and $v_{\rm sc}$ as above (thin dot-dashed curve), as well
as the \emph{lower limit} (`cold ambient plasma' regime; see equation)
for the Coulomb collision timescale $\tau_{\rm Coul} \simeq 0.01$\,Myr
of the background thermal gas (thin horizontal line); these are all
longer than $\tau_{\rm acc}$ and $\tau_{\rm rad}$ for $B \gtrsim$ a 
few $\mu$G. Figure\,\ref{fig:timescales} indicates therefore that,
assuming strong turbulent conditions, efficient acceleration
and confinement of the radiating electrons within compact and
substantially magnetized emission sites is plausible.

Importantly, for the particular conditions of the emission site
discussed here, namely a very inefficient particle escape when
compared with the acceleration and radiative losses timescales (and
also the lifetime of the system), as well as the stochastic nature of
the acceleration process, the resulting electron population is
expected to be of a `modified ultrarelativistic Maxwellian' form
$dN_e\!(\gamma)/d\gamma \propto \gamma^2
\exp\!\left[-\frac{1}{a}\left(\gamma/\gamma_{\rm eq}\right)^a\right]$
with the parameter $0< a < 2$ depending on particular cooling and
turbulence conditions, as discussed in \citet{Stawarz08}, i.e. to be
peaked (with a broader plateau) around the equilibrium energy
$\gamma_{\rm eq}$. The high-energy segment of the synchrotron
continuum produced by such electrons seems to be qualitatively
consistent with the observed spectral properties of the analyzed X-ray
spots \citep[see Figure\,12 in][]{Stawarz08}, while at lower
(radio--to--optical) frequencies very flat synchrotron continua would
be expected. For such an electron population, $\int \gamma^2
dN_e\!(\gamma) / \int \gamma \, dN_e\!(\gamma) \sim \gamma_{\rm eq}$,
and the total pressure of the discussed X-ray--emitting electron
population is expected to be rather tiny,
\begin{equation}
p_{e\pm} \simeq \frac{m_e c \, L_{\rm x}}{4 \sigma_T \, \gamma_{\rm eq} \, U_B \, V} \sim 2 \times 10^{-16} \, \bmg^{-3/2} \, R_{1.5}^{-3} \, {\rm erg\,cm^{-3}}
\end{equation}
where $\bmg \equiv B/$\,$\mu$G, so that it starts to exceed the thermal pressure of the surrounding medium $\langle p_g \rangle$ or the magnetic pressure $U_B$ within the emission site only for $R$ significantly smaller than hundreds of parsecs. This is the advantage of the synchrotron scenario over the previous two considered above. The interesting aspect of the model is the implied production of very high energy $\gamma$-rays ($\varepsilon_{\gamma} \gtrsim 10$\,TeV) by the synchrotron X-ray--emitting electrons via the accompanying IC/CMB process; the total \emph{diffuse} $\gamma$-ray flux thus expected for the entire giant halo can be evaluated very roughly as $F_{\rm \gamma,\,tot} \simeq (U_{\rm cmb} / U_B) \times F_{\rm X,\,tot} \sim 10^{-11}$\,\Su\ for the spots' magnetic field of the order of a few $\mu$G. 

\section{Summary and Conclusions}
\label{sec:summary}

In this paper we have discussed \suz\ observations of selected regions
within the Southern giant lobe of the \cen\ radio galaxy. In the
analysis we have focused first on distinct X-ray features detected with the
XIS instrument, at least some of which are likely associated with fine
structure of the lobe revealed by the most recent high-quality radio
intensity and polarization maps. We found that from the available
photon statistics, the spectral properties of the detected X-ray
features are equally consistent with with thermal emission from hot
gas with temperatures $kT \simeq 1-10$\,keV, or with power-law
radiation continua characterized by photon indices $\Gamma \simeq
2.0\pm 0.5$. The plasma parameters implied by these different models
favor a synchrotron origin of the analyzed X-ray spots, which indicates
that a very efficient acceleration of electrons up to $\gtrsim
10$\,TeV energies is taking place within the giant structure of \cen,
albeit only in isolated and compact regions associated with extended
and highly polarized radio filaments. We speculate that the magnetic
energy released locally at the sites of reconnecting magnetic
tubes/radio filaments loads turbulence into the system, and that this
turbulence is in turn responsible for the confinement and acceleration
of relativistic particles within distinct emission sites.
Alternatively, if most of the released magnetic energy goes into
heating of the plasma rather than acceleration of ultrarelativistic
particles, the thermal scenario, involving short-lived and highly
over-pressured gaseous clumps remains a valid possibility, suggesting
an efficient transfer of the magnetic energy to the internal energy of
the lobes' thermal plasma during the lifetime of the source.

Large-scale highly polarized radio filaments have been detected within
the lobes of several other famous radio galaxies
\citep{Fomalont89,Carilli91,Perley97,Swain98,Owen00}. Apparently
similar structures are observed in the Galactic Center region
\citep{Lang99,LaRosa00,Nord04,Zadeh04}. The origin and the exact
structure of such features is unclear, but tangled magnetic field
tubes are often invoked in this context
\citep[e.g.,][]{ONeill10}. The X-ray observations
reported here constitute an interesting insight into the problem,
suggesting in particular that, as in fact expected in the case of
tangled field tubes by analogy with solar phenomena, localized regions
of enhanced turbulence enable efficient cosmic ray acceleration and/or
plasma heating form a set of transient `hot spots' in compact sites
where separate filaments interact with each other or with
over-densities in the ambient plasma so that magnetic reconnection takes place.

While we have emphasized the possibility of reconnection acting as a source of the turbulence, there are other processes that could account for such a localized enhancement. If the \cen\ giant halo is indeed a source of very-high to ultra-high energy cosmic rays, as has been discussed in the literature \citep{Hardcastle09,OSullivan09,Peer12} prompted by the results of the P. Auger Observatory \citep[e.g.,][]{Moskalenko09,Nemmen10,Yuksel12}, the flux of these particles through such regions can in principle produce un-compensated currents, amplifying small-scale magnetic fields and turbulent motions, in a process analogous to that frequently considered in the context of supernova remnants \citep{Bell04}.

We have also presented a detailed analysis of the diffuse emission
component filling the whole FOV of the XIS instrument, which remains
after removing all the compact X-ray features. We found that at the
position of the \cen\ giant lobes probed with \suz\ a relatively
prominent soft X-ray excess, best described as thermal emission from
hot gas with the temperature of $kT \simeq 0.5$\,keV, seems to be
present. This excess, if related to the \cen\ system rather than some
large-scale fluctuation in the foreground Galactic Halo emission, may
be either due to thermal matter mixed with the non-thermal plasma of
the giant lobes, or due to a condensation of the hot intergalactic
medium around the edges of the expanding radio structure. The former
possibility would interestingly imply a number density of the X-ray
emitting gas of roughly $n_g \simeq 10^{-4}$\,cm$^{-3}$, as well as
almost exact pressure balance between the lobes thermal and
non-thermal contents (again in analogy with the interstellar medium
within the Galactic disk). These conclusions, which are still
tentative and somewhat preliminary, are in agreement with the most
recent analysis of the radio data presented by \citet{Shane12}. It is
important to emphasize that our finding regarding the plasma $\beta$
parameter being of the order of unity in the giant lobes of \cen\ may
be crucial for understanding the dynamics and evolution of the
reconnecting magnetic field in the system. \citet{Gourgouliatos12}
argued for example that the evolution of the expanding lobes
containing \emph{some} thermal plasma and \emph{some} magnetic flux
will always be accompanied by a decrease in the $\beta$ parameter,
leading at some point to the spontaneous formation of electric current
sheets and large-scale reconnection layers which are likely involved
in acceleration of the highest energy cosmic rays.
 
Regardless of the exact processes involved in the formation of the
observed X-ray sub-structure of the lobes, however, or of the origin
of the soft diffuse excess emission, the total X-ray emission of the
whole giant halo in \cen\ appears to be a complex mixture of
diffuse and compact thermal and non-thermal components. Only with the
broad energy coverage and the suitable combination of the sensitivity
and resolution of \suz\ has it been possible to disentangle these
components; otherwise they would form a single emission continuum. If our
preliminary conclusion that these various emission components are
indeed related to the \cen\ system is correct, there are important implications
for understanding the physics of radio-loud AGN: the
structure of the extended lobes in such systems may be highly
inhomogeneous and non-uniform, with magnetic reconnection and
turbulent acceleration processes continuously converting magnetic
energy to the internal energy of the plasma particles, leading to
possibly significant spatial and temporal variations in the plasma
$\beta$ parameter and pressure equilibrium conditions. Forthcoming
\suz\ pointings and the follow-up \textit{Chandra} observations of the
\cen\ giant lobes will shortly allow us to test the above conclusions.

Finally, we note that a possibly significant contribution from compact
features and thermal plasma to the total observed X-ray flux of the
extended lobes means that without detailed good-quality X-ray maps
enabling the various emission components to be disentangled, the ratio
of total X-ray and radio fluxes, typically used to infer
\emph{volume-averaged} magnetic field intensity (under the working
assumption that the inverse-Compton scattering of CMB photons by
radio-emitting electrons uniformly filling the lobe provides the
dominant contribution to the measured X-ray flux), should be used with
some caution \citep[see the related discussion in][]{Hardcastle07}. In
particular, departures from the energy equipartition often claimed
because of apparently high values of this ratio \citep[e.g.,][and
  references therein]{Kataoka05,Croston05,Isobe11}, may be partly
invalidated, and the lobes instead, of being dominated by the
relativistic particle pressure, may well be \emph{on average} very
close to energy equipartition, $\langle p_g \rangle \sim \langle
p_{e\pm} \rangle \sim \langle U_B \rangle$, as proposed in this paper for the \cen\ giant lobes.

\section*{Acknowledgments}

\L .~S. and M.~O. were supported by Polish NSC grant DEC-2012/04/A/ST9/00083.
S.~O'S acknowledges the support of the Australian Research Council through grant FL100100033.
Work by C.~C.~C.~at NRL is supported in part by NASA DPR S-15633-Y.
B.~R. has received funding from the European Research Council under the European Community's Seventh Framework Programme (FP7/2007-2013) / ERC grant agreement no. 247039.

\appendix

As a cross-check of the modeling of the diffuse emission component in
all four analyzed \suz\ pointings (see \S\,\ref{sec:diffuse}), we fit
the diffuse spectra obtained after removing all the point sources
present in the FOVs with the same model
\texttt{wabs*(APEC\_1+APEC\_2+PL)} including absorbed thermal
component (\texttt{APEC\_1}) representing Galactic Halo (GH) emission,
absorbed power-law component corresponding to the Cosmic X-ray
Background (CXB) radiation, and an additional absorbed thermal
component (\texttt{APEC\_2}) describing any residual thermal emission
related to the \cen\ system. 

We utilized the data in the $0.5-7$\,keV
photon energy range for XIS1 and $0.6-7$\,keV range for XIS0/3, and
applied C-statistics (preferred over the $\chi^2$-statistics in the
case of limited counts) in the \emph{simultaneous} spectral fitting.
The neutral hydrogen column density was fixed as before at the
Galactic value in the direction of the source, $N_{\rm H,\,Gal} = 0.7
\times 10^{21}$\,cm$^{-2}$. For the GH component, we fixed the
abundance at the solar value, and required the temperature and the
normalization to be the same in all four spectra. The photon index for
the CXB component was fixed as $\Gamma_{\rm CXB} = 1.41$ as before,
but the CXB normalization was allowed to vary between ``Lobe\,1\&2''
and ``Lobe\,3\&4'' spectra. This choice was dictated by the rather
different net exposures of the ``on'' and ``off'' pointings, and the
consequently different flux levels of the point sources removed, which
in particular affects the high-energy segment of the spectra. Finally,
the metallicity of the extra thermal component was fixed at a value of 
0.3 solar, but its temperature and normalization was allowed to vary 
between all pointings. The results of the fitting are summarized in 
Table\,\ref{tab:diff3}. As shown, an excess $\simeq 0.6$\,keV emission 
at the position of the lobes is recovered just as before, with the normalization 
of the \texttt{APEC\_2} component in the ``on'' pointings being about twice 
higher than in the ``off'' pointings (cf. Table\,\ref{tab:diff1}).

\begin{table*}[!h]
{\footnotesize
\noindent
{\caption[] {\label{tab:diff3} Alternative Modeling Results for the Diffuse Emission}}
\begin{center}
\begin{tabular}{ccccccccc}
\hline\hline
\\
region & $Z_{\rm GH}/Z_{\odot}$ & $kT_{\rm GH}$ & Norm$_{\rm GH}$ & $Z_{\rm ext}/Z_{\odot}$ & $kT_{\rm ext}$ & Norm$_{\rm ext}$ &  $\Gamma_{\rm CXB}$ & Norm$_{\rm CXB}$\\
(1) & (2) & (3) & (4) & (5) & (6) & (7) & (8) & (9) \\
\\
\hline
\\
Lobe\,1 (``on'') & $1.0^f$ & $0.16^{+0.01}_{-0.02}$ & $7.34^{+3.48}_{-1.52}$ & $0.3^f$ & $0.59\pm0.02$ & $3.60\pm0.17$  & $1.41^f$ & $1.04\pm0.02$ \\
Lobe\,2 (``on'') & '' & '' & '' & '' & $0.73\pm0.02$ & $3.15\pm0.15$  & '' &''  \\
Lobe\,3 (``off'') & '' & '' & '' & '' & $0.71\pm0.11$ & $1.43\pm0.27$  & '' &  $1.30^{+0.04}_{-0.03}$ \\
Lobe\,4 (``off'') & '' & '' & '' & '' & $1.59\pm0.36$ & $1.31\pm0.30$  &  '' &  ''   \\
\\
\hline\hline
\end{tabular}
\end{center}
(1) Region fitted; (2) fixed (``$f$'') abundance of the GH component (3) temperature of the GH component in the keV units; (4) \texttt{APEC} normalization of the GH component $\times 10^{-3}$; (5) fixed (``$f$'') abundance of the extra thermal component (6) temperature of the extra thermal component in the keV units; (7) \texttt{APEC} normalization of the extra thermal component $\times 10^{-3}$; (6) fixed (``$f$'') photon index of the CXB component; (9) normalization of the CXB component $\times 10^{-3}$.
}
\end{table*}

{}

\end{document}